\documentstyle[12pt,psfig]{article}
\def\beq{\begin{equation}}
\def\eeq{\end{equation}}
\def\bea{\begin{eqnarray}}
\def\eea{\end{eqnarray}}

\begin{document}
\thispagestyle{empty}
\noindent
\begin{flushright}
        OHSTPY-HEP-T-99-017\\
        NUHEP-TH-99-77\\
         CERN-TH/2000-03\\
        December 1999
\end{flushright}

\vspace{1cm}
\begin{center}

{\Large\bf
      Neutrino Oscillations in an SO(10) SUSY GUT with U(2)xU(1)$^n$ Family Symmetry}

  \vspace{1cm}

  {\Large\bf
    T. Bla\v{z}ek$^\dagger$$^\ddagger$,  S. Raby$^*$ and K. Tobe$^*$$^\#$}

    \bigskip
{\em $^\dagger$Department of Physics and Astronomy,
Northwestern University,
Evanston, IL 60208

$^*$Department of Physics,
The Ohio State University,
174 W. 18th Ave.,
Columbus, Ohio  43210

$^\#$Present address: Theory Division, CERN, CH-1211 Geneva 23, Switzerland}
\end{center}
\vspace{1cm}
\centerline{\bf Abstract}
\begin{quotation}
\noindent
  In a previous paper we analyzed fermion masses (focusing on neutrino masses and mixing angles) in an SO(10) SUSY GUT with U(2)$\times$U(1)$^n$ family symmetry.   The model is  ``natural" containing all operators in the Lagrangian consistent with the states and their charges.  With minimal family symmetry breaking vevs the model is also predictive giving a unique solution to atmospheric (with maximal $\nu_\mu \rightarrow \nu_\tau$ mixing) and solar (with SMA MSW $\nu_e \rightarrow \nu_s$ mixing) neutrino oscillations.   In this paper we analyze the case of general family breaking vevs.  We now find several new solutions for three, four and five neutrinos.   For three neutrinos we now obtain SMA MSW, LMA MSW or vacuum oscillation solutions for solar neutrinos.  In all three cases the atmospheric data is described by maximal $\nu_\mu \rightarrow \nu_\tau$ mixing.   In the four and five neutrino cases, in addition to fitting atmospheric and solar data as before, we are now able to fit LSND data.  All this is obtained with the additional parameters coming from the family symmetry breaking vevs; providing only minor changes in the charged fermion fits.
\end{quotation}

\vfill
$^\ddagger$ {\em
           On leave of absence from
           Faculty of Mathematics and Physics,
           Comenius Univ., Bratislava, Slovakia}

\section{Introduction}
Neutrino oscillations provide a window onto new physics beyond the standard model \footnote{In the Standard Model, the three active neutrino species (members of electroweak doublets) are massless.
As a consequence individual lepton number is conserved and neutrinos cannot
oscillate.}  and several experiments now provide evidence for neutrino oscillations. This includes data on solar neutrinos ~\cite{solar}, atmospheric neutrinos ~\cite{atmos} and the accelerator-based experiment, liquid scintillator neutrino detector [LSND]~\cite{LSND}.  These positive indications are constrained by null experiments such as Chooz~\cite{chooz} and Karmen~\cite{karmen}.   The data strongly suggests that neutrinos have small masses and non-vanishing mixing angles ~\cite{massmatrices}.  In the near future, many more experiments will test the hypothesis of neutrino masses~\cite{karmen}, \cite{sno} - \cite{kamland}.  Thus there is great excitement and anticipation in this field.

In a recent paper I~\cite{brt} we analyzed an $SO_{10}$ supersymmetric [SUSY] grand unified theory [GUT] with family symmetry $U_2 \times U_1^n$.  The theory was "natural," i.e. the Lagrangian was the most general consistent with the states and symmetries.   In addition, with {\em minimal family symmetry breaking vacuum expectation values [vevs]}, the number of arbitrary parameters in the effective low energy theory, below the GUT scale, was less than the number of observables.   Hence the theory was "predictive" and testable.  We analyzed the predictions for charged fermion masses and mixing angles using a global $\chi^2$ analysis~\cite{blazek,brt} finding excellent agreement with the data.   In the neutrino sector we obtained a {\em unique} solution to both atmospheric~\cite{atmos} and solar~\cite{solar} neutrino oscillation data.  This solution has three active and one sterile neutrino.  It has maximal $\nu_\mu \rightarrow \nu_\tau$ oscillations fitting atmospheric data and small mixing angle [SMA] Mikheyev, Smirnov, Wolfenstein [MSW]~\cite{msw}  $\nu_e \rightarrow \nu_s$ (where $s$ denotes $sterile$) oscillations for solar data, {\em without fine-tuning}.  We were however unable to simultaneously fit LSND~\cite{LSND}, even with four neutrinos.  In addition, we were unable to find a three neutrino solution to both atmospheric and solar neutrino data.   It is imperative to understand if these results are robust.   In particular, without a theory of family symmetry breaking we may consider more general family symmetry breaking vevs.\footnote{In fact, we noted in ~\cite{brt} that it is possible to obtain three neutrino solutions to atmospheric and solar data if we allow for {\em non-minimal} family symmetry breaking vevs.}   In this paper we allow for the most general family symmetry breaking vevs;  introducing two new complex parameters  $\kappa_{(1, 2)}$.  There are now more parameters for charged fermion masses and mixing angles than there are observables.  The new parameters have minor consequences for charged fermions (fits to $m_e$, $m_\mu$, and $V_{us}$, which
are all known to excellent accuracy, require them to remain small), but significant consequences for neutrinos.  In fact, with the additional parameters we are now able to obtain three possible three-neutrino solutions to atmospheric~\cite{atmos} and solar~\cite{solar} neutrino data.   With one or two sterile neutrinos we can also obtain solutions to atmospheric~\cite{atmos}, solar~\cite{solar} and LSND~\cite{LSND} data.

In section 2, we discuss the model and family symmetry breaking.  The model is an SO(10) [SUSY GUT] $\times$U(2)$\times$U(1)$^n$ [family symmetry] model.  It is a small variation of the theory introduced by Barbieri et al. [BHRR]~\cite{so10u2}  where the non-abelian family symmetry was introduced to provide a natural solution to flavor violation in SUSY theories~\cite{familysymmetry,u2symmetry,flavorviolation}.  In section 3, we present the general framework for neutrino masses and mixing angles.  In section 4, we describe the three neutrino solutions and in sections 5 and 6 we present the four and five neutrino solutions, respectively.  Our conclusions are in section 7.

\section{An SO(10)$\times$U(2)$\times$U(1)$\times \cdots$ model}

The three families of fermions are contained in $16_a, \, a = 1,2;$ and
$16_3$ where $a$ is a U(2) flavor index.  [Note U(2) = SU(2) $\times$ U(1)$'$ where the U(1)$'$ charge is +1 ($-1$) for each upper (lower) SU(2) index.]  At tree level, the third family of fermions couples to a $10$ of Higgs with coupling $\lambda \; 16_3 \; 10 \; 16_3$ in the superspace  potential.  The Higgs and $16_3$ have zero charge under the first two U(1)s, while $16_a$ has charge $-1$ and thus does not couple to the Higgs at tree level.~\footnote{There are in fact four additional U(1)s implicit in the superspace potential (eqn. \ref{eq:W}).  These are a Peccei-Quinn symmetry in which all 16s  have charge +1,  all $\overline{16}$s have charge $-1$, and 10 has charge $-2$; a flavon symmetry in which ($\phi^a, \; S^{a\, b}, \; A^{a \, b}$) and $M$ have charge +1 and $\bar \chi_b$ has charge $-1$; a symmetry in which $M', \; M''$ have charge +1 and $\bar \chi, \; \bar \chi^a$ have charge -1 and and an R symmetry in which all chiral superfields have charge +1. The family symmetries of the theory may be realized as either global or local symmetries.  For the purposes of this
paper, it is not necessary to specify which one.  However, if it is realized locally, as might be  expected from string theory, then not all of the U(1)s are anomaly free.   We would then need to specify the complete set of anomaly free U(1)s.}

Three superfields ($\phi^a, \; S^{a\, b} = S^{b\, a}, \; A^{a \, b} =
-A^{b\, a}$) are introduced to spontaneously break U(2)$\times$U(1) and to generate Yukawa terms giving mass to the first and second generations.  The fields ($\phi^a, \; S^{a\, b}, \; A^{a \, b}$) are SO(10) singlets with U(1) charges \{0, 1, 2\}, respectively.  The most general vacuum expectation values are given by
\bea
 \langle \phi^2 \rangle \:  \neq 0, &  \nonumber \\
 \langle S^{2 2} \rangle  \neq 0, &  
\langle S^{1 1} \rangle = \kappa_1 \, \langle S^{2 2} \rangle ,\;\;  
\langle S^{1 2} \rangle = \kappa_2 \, \langle S^{2 2} \rangle,  \label{eq:vevs} \\  \langle A^{1 2} \rangle  \neq 0, & \nonumber
\eea
where the constants $\kappa_1, \; \kappa_2$ are arbitrary.  The vevs ($\langle \phi^2 \rangle \sim \langle S^{2 2} \rangle    \sim \epsilon M_0^2/\langle 45 \rangle$) break U(2)$\times$U(1) to $\tilde U(1)$ and  ($A^{1 2} \sim \epsilon^\prime M_0$) completely.  In this model, second generation masses are of order $\epsilon$, while first generation masses are of order $\epsilon'^2/\epsilon$.  In paper I~\cite{brt} we analyzed this theory with minimal family breaking vevs ($\kappa_1 = \; \kappa_2 = 0$).   In this paper we show the effects of non-vanishing $\kappa_{(1. 2)}$.

The superspace potential for the charged fermion sector of this theory, including the heavy Froggatt-Nielsen states ~\cite{fn}, is given by
\begin{eqnarray}  W \supset & 16_3 \; 10\; 16_3 \;\; +\; \;  16_a \; 10 \;
\chi^a & \label{eq:W} \\
& + \;\; \bar \chi_a \; (M \; \chi^a \; +\; \phi^a \; \chi \; +\; S^{a\; b} \;
 \chi_b \; + \;  A^{a\; b} \; 16_b) & \nonumber \\
&+ \;\; \bar \chi^a \; (M' \; \chi_a \;\; + \;\; 45 \; 16_a) & \nonumber \\
& + \;\; \bar \chi \; (M'' \; \chi \;\; + \;\; 45 \; 16_3) & \nonumber
\end{eqnarray}
where 
\begin{equation}
M = M_0 (1\;\;+ \;\;  \alpha_0 \; X \;\; +\;\; \beta_0 \; Y).
\label{eq:M0}
\end{equation}
$X,\; Y$ are SO(10) breaking vevs in the adjoint representation with $X$
corresponding to the  U(1) in SO(10) which preserves SU(5),  $Y$ is
standard weak hypercharge and   $\alpha_0, \; \beta_0$ are arbitrary parameters.  The field $45$ is assumed to obtain a vev in the $B - L$ direction.  Note, this theory differs from [BHRR]~\cite{so10u2} only in that the fields $\phi^a$ and $S^{a\, b}$ are now SO(10) singlets (rather than SO(10) adjoints) and the SO(10) adjoint quantum numbers of these fields, necessary for acceptable masses and mixing angles, has been made explicit in the field $45$ with U(1) charge 1.\footnote{This change (see BHRR~\cite{so10u2}) is the reason for the  additional U(1)s.}  This theory thus requires much fewer SO(10) adjoints.   Moreover our neutrino mass solution depends heavily on this change.

The effective mass parameters $M_0, \; M', \; M''$ are SO(10) invariants.\footnote{The effective mass parameters represent vevs of SO(10) singlet chiral superfields.} The scales are assumed to satisfy  $M_0 \sim M' \sim  M'' \gg
\langle \phi^2 \rangle \sim \langle S^{2\, 2} \rangle \gg \langle A^{1\, 2}
\rangle $ where $M_0$ may be of order the GUT scale.
In the effective theory below $M_0$, the Froggatt-Nielsen states  \{$\chi,\,
\bar \chi,\ \chi^a,\, \bar \chi_a,\  \chi_a,\, \bar \chi^a$\}
may be integrated out, resulting in the effective Yukawa matrices for up
quarks, down quarks, charged leptons and the Dirac neutrino Yukawa matrix given by (see fig. 1)~\footnote{Note, we use the notation of BHRR~\cite{so10u2}.  The parameter $\rho$ vanishes in the limit $\beta_0 = 0$(see equations  ~\ref{eq:M0}, \ref{eq:yukawa}).    This is a consequence of the B-L vev in the 2 - 2 entry or the anti-symmetry of the coupling to $A^{a b}$ in the 1 - 2 element which is in conflict with the SU(5) invariance of $M$ in this limit which only allows for symmetric $u - \bar u$ couplings.}

\begin{eqnarray}
Y_u =&  \left(\begin{array}{ccc}  \kappa_1 \, \epsilon \, \rho & (\epsilon' + \kappa_2 \, \epsilon) \rho & 0 \\
             - (\epsilon' - \kappa_2 \, \epsilon) \rho &  \epsilon \rho & \epsilon r T_{\bar u}     \\
      0  & \epsilon r T_Q& 1 \end{array} \right) \; \lambda & \nonumber \\
Y_d =&  \left(\begin{array}{ccc}  \kappa_1 \, \epsilon & \epsilon' + \kappa_2 \, \epsilon & 0 \\
- (\epsilon' - \kappa_2 \, \epsilon)  &  \epsilon  & \epsilon r \sigma T_{\bar d}\\
0  & \epsilon r T_Q & 1 \end{array} \right) \; \lambda & \label{eq:yukawa} \\
Y_e =&  \left(\begin{array}{ccc}  3 \, \kappa_1 \, \epsilon & - (\epsilon' - 3 \, \kappa_2 \, \epsilon) & 0 \\
          \epsilon' + 3 \, \kappa_2 \, \epsilon &  3 \epsilon  & \epsilon r T_{\bar e} \\
  0  & \epsilon r \sigma T_L & 1 \end{array} \right) \; \lambda &
 \nonumber \\
Y_{\nu} =&  \left(\begin{array}{ccc}  3 \, \kappa_1 \, \epsilon \, \omega & - (\epsilon' - 3 \, \kappa_2 \, \epsilon) \, \omega & 0 \\
      (\epsilon' + 3 \, \kappa_2 \, \epsilon) \, \omega &  3 \epsilon  \omega & {1 \over 2} \epsilon r \omega 
T_{\bar \nu} \\
       0  & \epsilon r \sigma T_L& 1 \end{array} \right) \; \lambda &
 \nonumber
\end{eqnarray}
with  \begin{eqnarray} \omega = {2 \, \sigma \over 2 \, \sigma - 1}
\label{eq:omega} \end{eqnarray} and
\begin{eqnarray} T_f  = & (\rm Baryon\# - Lepton \#) &
\label{eq:Tf} \\
\rm for & f = \{Q,\bar u,\bar d, L,\bar e, \bar \nu\}.& \nonumber
\end{eqnarray}
\begin{figure}
\vspace{-1cm}
	\centerline{ \psfig{file=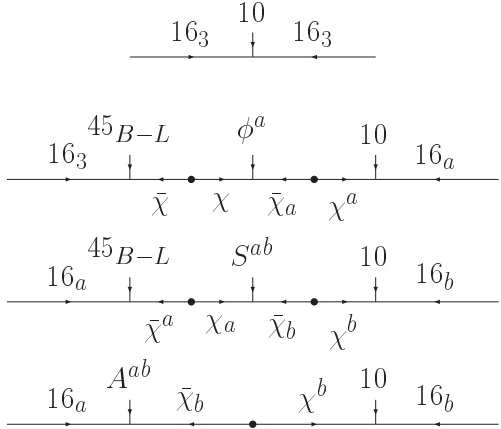,width=10cm,rheight=7cm}}
\caption{Diagrams generating the Yukawa matrices}
\end{figure}

In our notation, fermion doublets are on the left and singlets are on the
right.  Note, we have assumed that the Higgs doublets of the minimal supersymmetric standard model[MSSM] are contained in a single $10$ dimensional SO(10) multiplet.  Hence all the fits have large values of $\tan\beta$.~\footnote{Note, we could obtain small values of $\tan\beta$ in SO(10) at the cost of one new parameter. If the $10$ which couples to fermions mixes with other states then the Higgs field coupling to up and down quarks may have different effective couplings to matter, i.e. such that
$\lambda \;10 \supset \lambda \; H_u \; + \; \xi \; H_d$.  We could then
consider two limits ---  case (1)  $\lambda = \xi$ (no Higgs mixing) with large $\tan\beta$, and case (2)  $\lambda \gg \xi$ or small $\tan\beta$.  In paper I, we also considered case (2) and found no significant improvements in the fit.}

\subsection{Results for Charged Fermion Masses and Mixing Angles}

\protect
\begin{table}
\caption[8]{
{\bf Charged fermion masses and mixing angles} \\
   \mbox{Initial parameters: $\kappa_1 = \kappa_2 = 0$ }\ \ \
\ \

 (1/$\alpha_G, \, M_G, \, \epsilon_3$) = ($24.52, \, 3.03 \cdot 10^{16}$
GeV,$\,
-4.06$\%), \makebox[1.8em]{ }\\
 ($\lambda, \,$r$, \, \sigma, \, \epsilon, \, \rho, \, \epsilon^\prime$) =
($ 0.79, \,
12.4, \, 0.84, \, 0.011, \,  0.043,\,  0.0031$),\\
($\Phi_\sigma, \, \Phi_\epsilon, \, \Phi_\rho$) =  ($0.73, \, -1.21, \,
3.72$)rad,
\makebox[6.6em]{ }\\
($m_0, \, M_{1/2}, \, A_0, \, \mu(M_Z)$) = ($1000,\, 300, \, -1431, \,
110$) GeV,\\
($(m_{H_d}/m_0)^2, \, (m_{H_u}/m_0)^2, \, $tan$\beta$) = ($2.23,\, 1.66, \,
53.7$)
}
\label{t:fitk=0}
$$
\begin{array}{|l|c|l|}
\hline
{\rm Observable}  &{\rm Data}(\sigma) & Theory  \\
\mbox{ }   & {\rm (masses} & {\rm in\  \ GeV) }  \\
\hline
\;\;\;M_Z            &  91.187 \ (0.091)  &  91.17          \\
\;\;\;M_W             &  80.388 \ (0.080)    &  80.39       \\
\;\;\;G_{\mu}\cdot 10^5   &  1.1664 \ (0.0012) &  1.166     \\
\;\;\;\alpha_{EM}^{-1} &  137.04 \ (0.14)  &  137.0         \\
\;\;\;\alpha_s(M_Z)    &  0.1190 \ (0.003)   &  0.1174       \\
\;\;\;\rho_{new}\cdot 10^3  & -0.20 \ (1.1) & +0.314   \\
\hline
\;\;\;M_t              &  173.8 \ (5.0)   &  174.9 \\
\;\;\;m_b(M_b)          &    4.260 \ (0.11)  &    4.331                  \\
\;\;\;M_b - M_c        &    3.400 \ (0.2)   &    3.426                 \\
\;\;\;m_s              &  0.180 \ (0.050)   &  0.147          \\
\;\;\;m_d/m_s          &  0.050 \ (0.015)   &  0.0589        \\
\;\;\;Q^{-2}           &  0.00203 \ (0.00020)  &  0.00201                \\
\;\;\;M_{\tau}         &  1.777 \ (0.0018)   &  1.777         \\
\;\;\;M_{\mu}          & 0.10566 \ (0.00011)   & .1057           \\
\;\;\;M_e \cdot 10^3      &  0.5110 \ (0.00051) &  0.5110  \\
 \;\;\;V_{us}         &  0.2205 \ (0.0026)      &  0.2205        \\
\;\;\;V_{cb}         & 0.03920 \ (0.0030)      &  0.0403           \\
\;\;\;V_{ub}/V_{cb}    &  0.0800 \ (0.02)    &  0.0691                 \\
\;\;\;\hat B_K          &  0.860 \ (0.08)    &  0.870           \\
\hline
{B(b\!\rightarrow\! s \gamma)\!\cdot\!10^{4}}  &  3.000 \ (0.47) &  2.992  \\
\hline
  \multicolumn{2}{|l}{{\rm TOTAL}\;\;\;\; \chi^2}  2.26
            & \\
\hline
\end{array}
$$
\end{table}

We have performed a global $\chi^2$ analysis, incorporating two (one) loop
renormalization group[RG] running of dimensionless (dimensionful)
parameters from $M_G$ to $M_Z$ in the MSSM,  one loop radiative threshold corrections at $M_Z$, and 3 loop QCD (1 loop QED) RG running below $M_Z$.  Electroweak symmetry breaking is obtained self-consistently from the effective potential at one loop, with all one loop threshold corrections included. This analysis is performed using the code of Blazek et.al.~\cite{blazek}.~\footnote{We assume universal scalar mass $m_0$ for
squarks and sleptons at $M_G$.  We have not considered the flavor violating effects of U(2) breaking scalar masses in this paper.}  In this paper, we just
present the results for one set of soft SUSY breaking parameters $m_0, \;
M_{1/2}$ with all other parameters varied to obtain the best fit solution.  In the first two columns of table ~\ref{t:fitk=0} we give the 20 observables which enter the $\chi^2$ function, their experimental values and the uncertainty $\sigma$ (in parentheses).   In most cases $\sigma$ is determined by the 1 standard deviation experimental uncertainty, however in some cases the theoretical uncertainty ($\sim$ 0.1\%) inherent in our renormalization group running and one loop threshold corrections dominates.  Lastly, in contrast to paper I we include a 1999 updated value ~\cite{langacker1999} of $\rho_{new}$, the measure of SU(2) violation beyond the standard model.  This change substantially improves our global charged fermion fits.

There are 8 real Yukawa parameters and 5 complex phases. We take the complex phases to be $\Phi_\rho, \; \Phi_\epsilon, \; \Phi_\sigma, \; \Phi_{\kappa_1}$
and $\Phi_{\kappa_2}$.  With 13 fermion mass observables (charged fermion masses and mixing angles [$\hat{B}_K$ replacing $\epsilon_K$ as a ``measure of CP
violation"~\footnote{ The Jarlskog parameter $J = 
Im(V_{ud}V_{ub}^*V_{cb}V_{cd}^*)$ is a measure of CP violation.
We test $J$ by a comparison to the experimental value extracted from 
the well-known $K^0-\overline{K^0}$ mixing observable 
$\epsilon_K =  (2.26 \pm 0.02)\times10^{-3}$.  The largest uncertainty in such
a comparison, however, comes in the value of the QCD bag constant $\hat B_K$. 
 We thus exchange the Jarlskog parameter $J$ for $\hat B_K$ in the list 
of low-energy data we are fitting. Our theoretical value of $\hat B_K$ is 
defined as that value needed to agree with $\epsilon_K$
for a set of fermion masses and mixing angles derived from the
GUT-scale. We test this theoretical value against the ``experimental'' 
value of $\hat B_K$. This value, together with its error estimate, is 
obtained from recent lattice calculations~\cite{kilcup}.}  ]) we have enough parameters to {\em fit} the data.  In table ~\ref{t:fitk=0} we also show the fits obtained with $\kappa_1 = \kappa_2 = 0$ as a benchmark for the cases with non-zero $\kappa_{1,2}$ which follow. From table ~\ref{t:fitk=0} it is clear that this theory fits the low energy data quite well.~\footnote{Note, the strange quark mass $m_s(1 \rm GeV) \sim 150 \ \rm MeV$ is small, consistent with recent lattice results.}    

Finally, the squark, slepton, Higgs and gaugino spectrum of our theory is
consistent with all available data.  The lightest chargino and neutralino are
higgsino-like with the masses close to their respective experimental
limits. As an example of the additional predictions of this theory consider the CP violating mixing angles which may soon be observed at B factories.   For the selected fit with
$\kappa_1 = \kappa_2 = 0$ we find 
\begin{eqnarray}
(\sin 2\alpha, \, \sin 2\beta, \, \sin \gamma) = & (0.74, \, 0.54, \,
0.99)&
\end{eqnarray}  or equivalently the Wolfenstein parameters
\begin{eqnarray}
(\rho, \, \eta )    =      &( -0.04, \,      0.31)    &.
\end{eqnarray}

As an aside, we have also computed the SUSY contribution to the muon anomalous magnetic moment.  Our prediction for the selected SUSY point~\footnote{Although this result does depend on the particular point in SUSY parameter space we have selected, it is independent of the particular neutrino solution.  In addition, we have assumed universal masses for squarks and sleptons at the GUT scale.  Non-universal slepton masses can affect our result.} gives values for $a_\mu^{SUSY} \approx 40 \cdot 10^{-10}$, in good agreement with the latest preliminary data from the ongoing BNL experiment ~\cite{hertzog}.

In tables ~\ref{t:fit3nuSMAMSW} - \ref{t:fit4nu} we give results for non-zero
$\kappa_1, \; \kappa_2$.   These results have been obtained with a slightly different procedure than previously.  We have followed a multi step iterative procedure for finding ``good" fits to both charged fermion and neutrino data.  This is in lieu of combining the neutrino and charged fermion sectors into a single $\chi^2$ function and minimizing the total $\chi^2$ with respect to variations of all the parameters.   Let us now describe this procedure in more detail.

 In each case we select a pair of non-zero values for $\kappa_1$ and $\kappa_2$ and keep these two parameters fixed while we repeat the charged fermion analysis.  If we obtain a good fit, we use these as initial values for the analysis of the neutrino sector (discussed in the next section).  Then in the neutrino analysis we only vary those parameters not already included in the charged fermion analysis. If the resulting neutrino fit is not acceptable, we make a step in the ($\kappa_1, \; \kappa_2$) parameter space and start again with the charged fermion analysis.  We also found that we can improve the neutrino fit for fixed $\kappa_1$ and $\kappa_2$ if we return to the charged fermion analysis and carefully move one or more parameters entering the Yukawa matrices slightly away from their best fit value (watching so as not to incur large changes in the charged fermion contributions to $\chi^2$).  Thus our tables ~\ref{t:fit3nuSMAMSW} - \ref{t:fit4nu} do not show the absolute ``best" fits for fixed $\kappa_1$ and $\kappa_2$.  Following this procedure we focus independently on different neutrino solutions as indicated in the table captions.   Thus although the data in tables ~\ref{t:fit3nuSMAMSW} - \ref{t:fit4nu} do not seem much different, they do however represent significant changes in the neutrino sector; discussed in the next section.

\protect
\begin{table}
\caption[8]{
{\bf Charged fermion masses and mixing angles: 3 neutrino SMA MSW} \\
   \mbox{Initial parameters: $\kappa_1 = \kappa_2^2$, $|\kappa_2| = 0.028$ }\ \ \
\ \

 (1/$\alpha_G, \, M_G, \, \epsilon_3$) = ($24.52, \, 3.05 \cdot 10^{16}$
GeV,$\, -4.07$\%), \makebox[1.8em]{ }\\
 ($\lambda, \,$r$, \, \sigma, \, \epsilon, \, \rho, \, \epsilon^\prime$) =
($ 0.79, \,
12.3, \, -0.96, \, 0.010, \,  0.042,\,  0.0031$),\\
($\Phi_\sigma, \, \Phi_\epsilon, \, \Phi_\rho, \, \Phi_{\kappa_1}, \, \Phi_{\kappa_2}$) =  ($3.84, \, 0.0032, \,
5.02, \, -1.70, \, -0.85$)rad,
\makebox[3em]{ }\\
($m_0, \, M_{1/2}, \, A_0, \, \mu(M_Z)$) = ($1000,\, 300, \, -1438, \,
110$) GeV,\\
($(m_{H_d}/m_0)^2, \, (m_{H_u}/m_0)^2, \, $tan$\beta$) = ($2.22,\, 1.66, \,
53.7$)
}
\label{t:fit3nuSMAMSW}
$$
\begin{array}{|l|c|l|}
\hline
{\rm Observable}  &{\rm Data}(\sigma) & Theory  \\
\mbox{ }   & {\rm (masses} & {\rm in\  \ GeV) }  \\
\hline
\;\;\;M_Z            &  91.187 \ (0.091)  &  91.18          \\
\;\;\;M_W             &  80.388 \ (0.080)    &  80.40       \\
\;\;\;G_{\mu}\cdot 10^5   &  1.1664 \ (0.0012) &  1.166     \\
\;\;\;\alpha_{EM}^{-1} &  137.04 \ (0.14)  &  137.0         \\
\;\;\;\alpha_s(M_Z)    &  0.1190 \ (0.003)   &  0.1174       \\
\;\;\;\rho_{new}\cdot 10^3  & -0.20 \ (1.1) & +0.322   \\
\hline
\;\;\;M_t              &  173.8 \ (5.0)   &  175.0       \\
\;\;\;m_b(M_b)          &    4.260 \ (0.11)  &    4.326                  \\
\;\;\;M_b - M_c        &    3.400 \ (0.2)   &    3.432                 \\
\;\;\;m_s              &  0.180 \ (0.050)   &  0.146          \\
\;\;\;m_d/m_s          &  0.050 \ (0.015)   &  0.0585        \\
\;\;\;Q^{-2}           &  0.00203 \ (0.00020)  &  0.00201                \\
\;\;\;M_{\tau}         &  1.777 \ (0.0018)   &  1.776         \\
\;\;\;M_{\mu}          & 0.10566 \ (0.00011)   & .1057           \\
\;\;\;M_e \cdot 10^3      &  0.5110 \ (0.00051) &  0.5110  \\
 \;\;\;V_{us}         &  0.2205 \ (0.0026)      &  0.2206        \\
\;\;\;V_{cb}         & 0.03920 \ (0.0030)      &  0.0402           \\
\;\;\;V_{ub}/V_{cb}    &  0.0800 \ (0.02)    &  0.0702                 \\
\;\;\;\hat B_K          &  0.860 \ (0.08)    &  0.8691           \\
\hline
{B(b\!\rightarrow\! s \gamma)\!\cdot\!10^{4}}  &  3.000 \ (0.47) &  2.958  \\
\hline
  \multicolumn{2}{|l}{{\rm TOTAL}\;\;\;\; \chi^2}  2.48
            & \\
\hline
\end{array}
$$
\end{table}

\protect
\begin{table}
\caption[8]{
{\bf Charged fermion masses and mixing angles: 3 neutrino LMA MSW} \\
   \mbox{Initial parameters: $|\kappa_1| = 0.055$, $|\kappa_2| = 0.31$ }\ \ \
\ \

 (1/$\alpha_G, \, M_G, \, \epsilon_3$) = ($24.52, \, 3.05 \cdot 10^{16}$
GeV,$\, -4.08$\%), \makebox[1.8em]{ }\\
 ($\lambda, \,$r$, \, \sigma, \, \epsilon, \, \rho, \, \epsilon^\prime$) =
($ 0.79, \,
14.3, \, -1.13, \, 0.009, \,  0.045,\,  0.0028$),\\
($\Phi_\sigma, \, \Phi_\epsilon, \, \Phi_\rho, \, \Phi_{\kappa_1}, \, \Phi_{\kappa_2}$) =  ($3.82, \, -0.69, \,
4.83, \, 4.07, \, -1.14$)rad,
\makebox[3em]{ }\\
($m_0, \, M_{1/2}, \, A_0, \, \mu(M_Z)$) = ($1000,\, 300, \, -1444, \,
110$) GeV,\\
($(m_{H_d}/m_0)^2, \, (m_{H_u}/m_0)^2, \, $tan$\beta$) = ($2.22,\, 1.66, \,
53.7$)
}
\label{t:fit3nuLMAMSW}
$$
\begin{array}{|l|c|l|}
\hline
{\rm Observable}  &{\rm Data}(\sigma) & Theory  \\
\mbox{ }   & {\rm (masses} & {\rm in\  \ GeV) }  \\
\hline
\;\;\;M_Z            &  91.187 \ (0.091)  &  91.18          \\
\;\;\;M_W             &  80.388 \ (0.080)    &  80.40       \\
\;\;\;G_{\mu}\cdot 10^5   &  1.1664 \ (0.0012) &  1.166     \\
\;\;\;\alpha_{EM}^{-1} &  137.04 \ (0.14)  &  137.0         \\
\;\;\;\alpha_s(M_Z)    &  0.1190 \ (0.003)   &  0.1174       \\
\;\;\;\rho_{new}\cdot 10^3  & -0.20 \ (1.1) & +0.322   \\
\hline
\;\;\;M_t              &  173.8 \ (5.0)   &  174.9\\
\;\;\;m_b(M_b)          &    4.260 \ (0.11)  &    4.323                  \\
\;\;\;M_b - M_c        &    3.400 \ (0.2)   &    3.433                 \\
\;\;\;m_s              &  0.180 \ (0.050)   &  0.138          \\
\;\;\;m_d/m_s          &  0.050 \ (0.015)   &  0.0664        \\
\;\;\;Q^{-2}           &  0.00203 \ (0.00020)  &  0.00202                \\
\;\;\;M_{\tau}         &  1.777 \ (0.0018)   &  1.776         \\
\;\;\;M_{\mu}          & 0.10566 \ (0.00011)   & .1057           \\
\;\;\;M_e \cdot 10^3      &  0.5110 \ (0.00051) &  0.5110  \\
 \;\;\;V_{us}         &  0.2205 \ (0.0026)      &  0.2204        \\
\;\;\;V_{cb}         & 0.03920 \ (0.0030)      &  0.0409           \\
\;\;\;V_{ub}/V_{cb}    &  0.0800 \ (0.02)    &  0.0782                 \\
\;\;\;\hat B_K          &  0.860 \ (0.08)    &  0.8682           \\
\hline
{B(b\!\rightarrow\! s \gamma)\!\cdot\!10^{4}}  &  3.000 \ (0.47) &  2.999  \\
\hline
  \multicolumn{2}{|l}{{\rm TOTAL}\;\;\;\; \chi^2}  3.99            & \\
\hline
\end{array}
$$
\end{table}

\protect
\begin{table}
\caption[8]{
{\bf Charged fermion masses and mixing angles: 3 neutrino Vacuum} \\
   \mbox{Initial parameters: $|\kappa_1| = 0.004$, $|\kappa_2| = 0.025$}\ \ \
\ \

 (1/$\alpha_G, \, M_G, \, \epsilon_3$) = ($24.52, \, 3.05 \cdot 10^{16}$
GeV,$\, -4.16$\%), \makebox[1.8em]{ }\\
 ($\lambda, \,$r$, \, \sigma, \, \epsilon, \, \rho, \, \epsilon^\prime$) =
($ 0.80, \,
15.6, \, -0.35, \, 0.013, \,  0.041,\,  0.0035$),\\
($\Phi_\sigma, \, \Phi_\epsilon, \, \Phi_\rho, \, \Phi_{\kappa_1}, \, \Phi_{\kappa_2}$) =  ($3.00, \, -0.65, \,
4.41, \, 3.74, \, -0.052$)rad,
\makebox[3em]{ }\\
($m_0, \, M_{1/2}, \, A_0, \, \mu(M_Z)$) = ($1000,\, 300, \, -1433, \,
110$) GeV,\\
($(m_{H_d}/m_0)^2, \, (m_{H_u}/m_0)^2, \, $tan$\beta$) = ($2.22,\, 1.66, \,
53.7$)
}
\label{t:fit3nuVacuum}
$$
\begin{array}{|l|c|l|}
\hline
{\rm Observable}  &{\rm Data}(\sigma) & Theory  \\
\mbox{ }   & {\rm (masses} & {\rm in\  \ GeV) }  \\
\hline
\;\;\;M_Z            &  91.187 \ (0.091)  &  91.18          \\
\;\;\;M_W             &  80.388 \ (0.080)    &  80.40       \\
\;\;\;G_{\mu}\cdot 10^5   &  1.1664 \ (0.0012) &  1.166     \\
\;\;\;\alpha_{EM}^{-1} &  137.04 \ (0.14)  &  137.0         \\
\;\;\;\alpha_s(M_Z)    &  0.1190 \ (0.003)   &  0.1171       \\
\;\;\;\rho_{new}\cdot 10^3  & -0.20 \ (1.1) & +0.322   \\
\hline
\;\;\;M_t              &  173.8 \ (5.0)   &  175.0\\
\;\;\;m_b(M_b)          &    4.260 \ (0.11)  &    4.324                  \\
\;\;\;M_b - M_c        &    3.400 \ (0.2)   &    3.405                 \\
\;\;\;m_s              &  0.180 \ (0.050)   &  0.170          \\
\;\;\;m_d/m_s          &  0.050 \ (0.015)   &  0.0548        \\
\;\;\;Q^{-2}           &  0.00203 \ (0.00020)  &  0.00202                \\
\;\;\;M_{\tau}         &  1.777 \ (0.0018)   &  1.776         \\
\;\;\;M_{\mu}          & 0.10566 \ (0.00011)   & .1057           \\
\;\;\;M_e \cdot 10^3      &  0.5110 \ (0.00051) &  0.5110  \\
 \;\;\;V_{us}         &  0.2205 \ (0.0026)      &  0.2205        \\
\;\;\;V_{cb}         & 0.03920 \ (0.0030)      &  0.0392           \\
\;\;\;V_{ub}/V_{cb}    &  0.0800 \ (0.02)    &  0.0758                 \\
\;\;\;\hat B_K          &  0.860 \ (0.08)    &  0.8604           \\
\hline
{B(b\!\rightarrow\! s \gamma)\!\cdot\!10^{4}}  &  3.000 \ (0.47) &  2.938  \\
\hline
  \multicolumn{2}{|l}{{\rm TOTAL}\;\;\;\; \chi^2}  1.47            & \\
\hline
\end{array}
$$
\end{table}

\protect
\begin{table}
\caption[8]{
{\bf Charged fermion masses and mixing angles: 4 neutrino SMA MSW + LSND} \\
   \mbox{Initial parameters: $|\kappa_1| = 0.0001$, $|\kappa_2| = 0.002$ }\ \ \
\ \

 (1/$\alpha_G, \, M_G, \, \epsilon_3$) = ($24.50, \, 3.07 \cdot 10^{16}$
GeV,$\, -4.14$\%), \makebox[1.8em]{ }\\
 ($\lambda, \,$r$, \, \sigma, \, \epsilon, \, \rho, \, \epsilon^\prime$) =
($ 0.75, \,
12.4, \, -0.76, \, 0.011, \,  0.044,\,  0.0032$),\\
($\Phi_\sigma, \, \Phi_\epsilon, \, \Phi_\rho, \, \Phi_{\kappa_1}, \, \Phi_{\kappa_2}$) =  ($3.87, \, -0.95, \,
3.97, \, 4.81, \, 1.13$)rad,
\makebox[3em]{ }\\
($m_0, \, M_{1/2}, \, A_0, \, \mu(M_Z)$) = ($1000,\, 300, \, -1459, \,
110$) GeV,\\
($(m_{H_d}/m_0)^2, \, (m_{H_u}/m_0)^2, \, $tan$\beta$) = ($2.19,\, 1.65, \,
53.0$)
}
\label{t:fit4nu}
$$
\begin{array}{|l|c|l|}
\hline
{\rm Observable}  &{\rm Data}(\sigma) & Theory  \\
\mbox{ }   & {\rm (masses} & {\rm in\  \ GeV) }  \\
\hline
\;\;\;M_Z            &  91.187 \ (0.091)  &  91.18          \\
\;\;\;M_W             &  80.388 \ (0.080)    &  80.40       \\
\;\;\;G_{\mu}\cdot 10^5   &  1.1664 \ (0.0012) &  1.166     \\
\;\;\;\alpha_{EM}^{-1} &  137.04 \ (0.14)  &  137.0         \\
\;\;\;\alpha_s(M_Z)    &  0.1190 \ (0.003)   &  0.1173       \\
\;\;\;\rho_{new}\cdot 10^3  & -0.20 \ (1.1) & +0.318   \\
\hline
\;\;\;M_t              &  173.8 \ (5.0)   &  173.5\\
\;\;\;m_b(M_b)          &    4.260 \ (0.11)  &    4.341                  \\
\;\;\;M_b - M_c        &    3.400 \ (0.2)   &    3.422                 \\
\;\;\;m_s              &  0.180 \ (0.050)   &  0.148          \\
\;\;\;m_d/m_s          &  0.050 \ (0.015)   &  0.0591        \\
\;\;\;Q^{-2}           &  0.00203 \ (0.00020)  &  0.00201                \\
\;\;\;M_{\tau}         &  1.777 \ (0.0018)   &  1.776         \\
\;\;\;M_{\mu}          & 0.10566 \ (0.00011)   & .1057           \\
\;\;\;M_e \cdot 10^3      &  0.5110 \ (0.00051) &  0.5110  \\
 \;\;\;V_{us}         &  0.2205 \ (0.0026)      &  0.2205        \\
\;\;\;V_{cb}         & 0.03920 \ (0.0030)      &  0.0402           \\
\;\;\;V_{ub}/V_{cb}    &  0.0800 \ (0.02)    &  0.0699                 \\
\;\;\;\hat B_K          &  0.860 \ (0.08)    &  0.8696           \\
\hline
{B(b\!\rightarrow\! s \gamma)\!\cdot\!10^{4}}  &  3.000 \ (0.47) &  3.007  \\
\hline
  \multicolumn{2}{|l}{{\rm TOTAL}\;\;\;\; \chi^2}  2.94            & \\
\hline
\end{array}
$$
\end{table}

\protect
\begin{table}
\caption[8]{
{\bf Charged fermion masses and mixing angles: 5 neutrino SMA MSW + LSND} \\
   \mbox{Initial parameters: $|\kappa_1| = |\kappa_2|^2$, $|\kappa_2| = 0.032$ }\ \ \
\ \

 (1/$\alpha_G, \, M_G, \, \epsilon_3$) = ($24.52, \, 3.06 \cdot 10^{16}$
GeV,$\, -4.09$\%), \makebox[1.8em]{ }\\
 ($\lambda, \,$r$, \, \sigma, \, \epsilon, \, \rho, \, \epsilon^\prime$) =
($ 0.79, \, 12.2, \, -0.94, \, 0.011, \,  0.042,\,  0.0031$),\\
($\Phi_\sigma, \, \Phi_\epsilon, \, \Phi_\rho, \, \Phi_{\kappa_1}, \, \Phi_{\kappa_2}$) =  ($3.84, \, 0.07, \,
5.03, \, -2.49, \, -1.19$)rad,
\makebox[3em]{ }\\
($m_0, \, M_{1/2}, \, A_0, \, \mu(M_Z)$) = ($1000,\, 300, \, -1438, \,
110$) GeV,\\
($(m_{H_d}/m_0)^2, \, (m_{H_u}/m_0)^2, \, $tan$\beta$) = ($2.22,\, 1.66, \,
53.7$)
}
\label{t:fit5nu}
$$
\begin{array}{|l|c|l|}
\hline
{\rm Observable}  &{\rm Data}(\sigma) & Theory  \\
\mbox{ }   & {\rm (masses} & {\rm in\  \ GeV) }  \\
\hline
\;\;\;M_Z            &  91.187 \ (0.091)  &  91.17          \\
\;\;\;M_W             &  80.388 \ (0.080)    &  80.40       \\
\;\;\;G_{\mu}\cdot 10^5   &  1.1664 \ (0.0012) &  1.166     \\
\;\;\;\alpha_{EM}^{-1} &  137.04 \ (0.14)  &  137.0         \\
\;\;\;\alpha_s(M_Z)    &  0.1190 \ (0.003)   &  0.1174       \\
\;\;\;\rho_{new}\cdot 10^3  & -0.20 \ (1.1) & +0.322   \\
\hline
\;\;\;M_t              &  173.8 \ (5.0)   &  175.0\\
\;\;\;m_b(M_b)          &    4.260 \ (0.11)  &    4.328                  \\
\;\;\;M_b - M_c        &    3.400 \ (0.2)   &    3.426                 \\
\;\;\;m_s              &  0.180 \ (0.050)   &  0.148          \\
\;\;\;m_d/m_s          &  0.050 \ (0.015)   &  0.0588        \\
\;\;\;Q^{-2}           &  0.00203 \ (0.00020)  &  0.00201                \\
\;\;\;M_{\tau}         &  1.777 \ (0.0018)   &  1.777         \\
\;\;\;M_{\mu}          & 0.10566 \ (0.00011)   & .1057           \\
\;\;\;M_e \cdot 10^3      &  0.5110 \ (0.00051) &  0.5110  \\
 \;\;\;V_{us}         &  0.2205 \ (0.0026)      &  0.2205        \\
\;\;\;V_{cb}         & 0.03920 \ (0.0030)      &  0.0401           \\
\;\;\;V_{ub}/V_{cb}    &  0.0800 \ (0.02)    &  0.0701                 \\
\;\;\;\hat B_K          &  0.860 \ (0.08)    &  0.8686           \\
\hline
{B(b\!\rightarrow\! s \gamma)\!\cdot\!10^{4}}  &  3.000 \ (0.47) &  2.983  \\
\hline
  \multicolumn{2}{|l}{{\rm TOTAL}\;\;\;\; \chi^2}  2.12            & \\
\hline
\end{array}
$$
\end{table}

Before we conclude this section, let us consider one test (in the charged fermion sector) which may be able to distinguish among these different neutrino cases.   The unitarity angles 
$(\sin 2\alpha, \, \sin 2\beta, \, \sin \gamma)$ or equivalently the Wolfenstein parameters $(\rho, \; \eta)$ in some cases have significant corrections depending on the neutrino solution (see table ~\ref{t:wolfparameters}).   In particular, for larger values of $\kappa_1, \; \kappa_2$  we obtain significantly larger negative values of $\rho$; in particular consider $\rho = -0.24$ for the 3 $\nu$ LMA MSW solution.  This may be severely constrained by $B -\bar B$ mixing data.  However in order to determine whether this is consistent with present data we must first extend our numerical analysis to include this process.  We will look at this in a future paper~\cite{bdmrt}. 

\protect
\begin{table}
\caption[8]{
{\bf Unitarity triangle angles and Wolfenstein parameters for the different neutrino fits with non zero $\kappa_1, \; \kappa_2$.} \\
}
\label{t:wolfparameters}
$$
\begin{array}{|l|l|l|l|}
\hline
{\rm Neutrino \;\; fit} &{\rm Values \;\; of} \; \kappa_1, \; \kappa_2  & (\sin 2\alpha, \; \sin 2\beta, \; \sin \gamma) & (\rho, \; \eta)  \\
\hline
3 \nu \;\;\; {\rm SMA \, MSW} & \kappa_1 = \kappa_2^2, \;|\kappa_2| = 0.028& (0.92, \;  0.50, \; 0.95)  &  (-0.10, \;     0.30) \\
\hline 
3 \nu \;\;\; {\rm LMA \, MSW} & |\kappa_1| = 0.055,\; |\kappa_2| = 0.31 & (0.94, \; 0.39, \; 0.73) & ( -0.24, \;      0.26)    \\
\hline 
3 \nu \;\;\; {\rm Vacuum} & |\kappa_1| = 0.004,\; |\kappa_2| = 0.025& (0.86, \; 0.56, \; 0.97)& ( -0.08, \;      0.33)    \\
\hline 
4 \nu  {\rm SMA \, MSW + LSND} &|\kappa_1| = 0.0001,\; |\kappa_2| = 0.002 & (0.75, \; 0.54, \; 0.99) & ( -0.04, \;      0.31)   \\
\hline
5 \nu  {\rm SMA \, MSW + LSND} &|\kappa_1| = |\kappa_2|^2,\; |\kappa_2| = 0.032 & (0.88, \; 0.51, \; 0.96) & ( -0.09, \;      0.31)   \\
\hline
\end{array}
$$
\end{table}

\section{Neutrino Masses and Mixing Angles}

The parameters in the  Dirac Yukawa matrix for neutrinos (eqn.
~\ref{eq:yukawa}) mixing
$\nu - \bar \nu$ are now fixed.  Of course, neutrino masses are much too
large and we need to invoke the GRSY~\cite{grsy} see-saw mechanism.

Since the {\bf 16} of SO(10) contains the  ``right-handed" neutrinos
$\bar \nu$, one possibility is to obtain  $\bar \nu - \bar \nu$ Majorana masses
via higher dimension operators of the form~\footnote{This possibility
has been considered in the paper by Carone and Hall~\cite{u2neutrino}.}
\begin{equation}
{1 \over M} \ \overline{16} \ 16_3 \ \overline{16} \ 16_3   , 
{1 \over M^2} \ \overline{16} \ 16_3 \ \overline{16} \ 16_a \ \phi^{a}  ,
 {1 \over M^2} \ \overline{16} \ 16_a \ \overline{16} \ 16_b \
S^{a \, b} .
 \end{equation}

The second possibility, which we follow, is to introduce SO(10) singlet
fields $N$ and obtain effective mass terms $\bar \nu - N$ and $N - N$
using only dimension four operators in the superspace potential.  To do this,
we add three new SO(10) singlets
\{$N_a,\; a = 1,2;\;\; N_3$\}  with U(1) charges \{  $-1/2$,\  +1/2 \}.
These then contribute to the superspace potential
\begin{equation}
W \supset   \overline{16} \; (N_a \; \chi^a \;\; + \;\; N_3 \; 16_3) 
 + {1 \over 2} \ N_a \; N_b \; S^{a \; b} \;\; + \;\; N_a \; N_3 \; \phi^a
\end{equation}
where the field $\overline{16}$ with U(1) charge  $-1/2$ is assumed to get a
vev in the ``right-handed" neutrino direction. Note, this vev is
also needed to break the rank of SO(10).

Finally we allow for the possibility of adding a U(2) doublet of SO(10)
singlets
$\bar N^a$ or a U(2) singlet $\bar N^3$.  They enter the superspace
potential as
follows --
\begin{eqnarray}
W \supset &  \mu' \; N_a \; \bar N^a \;\; + \;\;\mu_3 \;  N_3 \bar N^3 &
\label{eq:mu'}
\end{eqnarray}
The dimensionful parameters $\mu', \; \mu_3$ are assumed to be of order the
weak
scale.  The notation is suggestive of the similarity between these terms and
the $\mu$ term in the Higgs sector. In both cases, we are adding
supersymmetric mass
terms and in both cases, we need some mechanism to keep these dimensionful
parameters
small compared to the Planck scale.

We define the 3 $\times $ 3 matrix
\begin{eqnarray}
\tilde \mu = & \left( \begin{array}{ccc}  \mu' & 0 & 0 \\
                               0 & \mu' & 0 \\
0 &  0  & \mu_3 \end{array}\right) &
\label{eq:mu}
\end{eqnarray}
The matrix $\tilde \mu$ determines the number of {\em coupled} sterile
neutrinos, i.e.
there are 4 cases labeled by the number of neutrinos ($N_\nu = 3, 4, 5, 6$):
\begin{itemize}
\item ($N_\nu = 3$) \hspace{.1cm} 3 active \hspace{.3cm} ($\mu' = \mu_3 =
0$);
\item ($N_\nu = 4$) \hspace{.1cm} 3 active + 1 sterile

\hspace{3cm} ($\mu' = 0;\; \mu_3 \neq 0$);
\item ($N_\nu = 5$) \hspace{.1cm} 3 active + 2 sterile

\hspace{3cm} ($\mu' \neq 0;\; \mu_3 = 0$);
\item ($N_\nu = 6$) \hspace{.1cm} 3 active + 3 sterile

\hspace{3cm} ($\mu' \neq 0;\; \mu_3 \neq 0$);
\end{itemize}
In this paper we consider the cases  $N_\nu = 3$, 4 and 5.

The generalized neutrino mass matrix is then given by~\footnote{This
is similar to the double see-saw mechanism suggested by Mohapatra and
Valle~\cite{mohapatra}.}

\begin{eqnarray}
& ( \begin{array}{cccc}\; \nu & \;\; \bar N & \;\; \bar \nu & \;\;  N
\end{array})  &
\nonumber\\  &  \left( \begin{array}{cccc}  0 & 0 & m & 0  \\
                     0 & 0 & 0 & \tilde \mu^T \\
                     m^T & 0 & 0 & V \\
                     0 & \tilde \mu & V^T & M_N  \end{array} \right) &
\end{eqnarray}
where  \begin{eqnarray} m = & Y_{\nu}\; \langle H_u^0 \rangle
&= \; Y_{\nu}\; {v \over\sqrt{2}}\; \sin\beta \end{eqnarray} and
\begin{eqnarray}
V = & \left( \begin{array}{ccc}  3 \, \kappa_1 \, \epsilon \, V_{16} & (\epsilon' + 3 \, \kappa_2 \, \epsilon)\, V_{16} & 0 \\
                                - (\epsilon' - 3 \, \kappa_2 \, \epsilon)\,  V_{16} & 3 \epsilon V_{16} & 0 \\
                                 0  & r \, \epsilon \, (1 - \sigma) \,
T_{\bar \nu} V_{16}  &
V'_{16}
\end{array}\right) &
\\
  M_N = & \left( \begin{array}{ccc}  \kappa_1 \, S & \kappa_2 \, S & 0 \\
                                \kappa_2 \, S & S & \phi \\
                                 0  & \phi  &  0 \end{array}\right) &
\nonumber
\end{eqnarray}
$V_{16},\; V'_{16}$ are proportional to the vev of $\overline{16}$
(with different
implicit Yukawa couplings) and $S, \; \phi$ are up to couplings the vevs of $
S^{22}, \; \phi^2$, respectively.

Since both $V$ and $M_N$ are of order the GUT scale, the states $\bar \nu,
\; N$
may be integrated out of the effective low energy theory.  In this case, the
effective neutrino mass matrix is given (at $M_G$) by~\footnote{In fact, at the
GUT scale $M_G$ we define an effective dimension 5 supersymmetric neutrino mass
operator where the Higgs vev is replaced by the Higgs doublet  H$_u$ coupled
to the entire lepton doublet.  This effective operator is then renormalized
using one-loop
renormalization group equations to $M_Z$.  It is only then that $H_u$ is
replaced by its
vev.} (the matrix is written in the ($\nu, \ \bar N$) {\em flavor} basis
where  charged lepton masses are diagonal)

\begin{equation}
m_\nu =   \tilde U_e^T \; \left( \begin{array}{cc}
m\;(V^T)^{-1}\;M_N\; V^{-1}\; m^T &  - m \;(V^T)^{-1}\; \tilde \mu\\
                    - {\tilde \mu}^T \; V^{-1}\; m^T & 0  \end{array}
\right) \; \tilde
U_e   \label{eq:generalmnu} 
\end{equation}
with
\begin{eqnarray} \tilde U_e = & \left(\begin{array}{cc} U_e & 0 \\
                                               0 & 1 \end{array}\right) & \\
e_0 =  U_e \; e \;\; ; &  \nu_0 = U_e \; \nu  &  \nonumber
\end{eqnarray}
 $U_e$ is the $3\times3$ unitary matrix for left-handed leptons needed to
diagonalize $Y_e$ (eqn. \ref{eq:yukawa}) and $e_0,\; \nu_0 \; (e, \; \nu)$
represent the
three families of left-handed leptons in the charged-weak ( -mass)
eigenstate basis.

The neutrino mass matrix is diagonalized by a unitary matrix $U =
U_{\alpha\, i}$;
\begin{eqnarray}
m^{diag}_{\nu} = & U^T \; m_{\nu} \; U & \label{eq:mnu}
\end{eqnarray}
where $\alpha= \{\nu_e ,\; \nu_\mu ,\; \nu_\tau ,\; \nu_{s_1}, \;
\nu_{s_2}, \; \nu_{s_3} \}$ is the flavor index and
$i = \{ 1, \cdots, 6\}$ is the neutrino mass eigenstate index.
$U_{\alpha\, i}$   is
observable in neutrino oscillation experiments.    In particular,  the
probability
for the flavor state $\nu_\alpha$ with energy $E$ to oscillate into
$\nu_\beta$ after
travelling a distance $L$ is given by
\begin{equation}
P(\nu_\alpha \rightarrow \nu_\beta)  =  \delta_{\alpha \beta}
- 4\sum_{k\; <\, j} U_{\alpha \, k} \ U^*_{\beta \, k} \
U^*_{\alpha \, j} \ U_{\beta \, j} \ \sin^2\Delta_{j\, k}  
\end{equation}
where $\Delta_{j\,k} =  {\delta m^2_{j k} \ L \over 4 E}$ and
$\delta m^2_{j k} = m^2_j - m^2_k$.

For N$_\nu \leq 4$ we have
\begin{equation} 
m_{\nu} =   m' \; \tilde U_e^T \; \left(
\begin{array}{cccc}  \kappa_1 \, \omega \, \zeta & \kappa_2 \, \omega \, \zeta & \frac{\kappa_1 \, \epsilon \, \epsilon^\prime \, r \, \sigma}{\bar \epsilon^2} \, \zeta & 0 \\
    \kappa_2 \, \omega \, \zeta  & b & C_{2 3} &  - u \, c \\
     \frac{\kappa_1 \, \epsilon \, \epsilon^\prime \, r \, \sigma}{\bar \epsilon^2} \, \zeta & C_{2 3} & C_{3 3} & - f \, c \\
0 &  - u \, c &  - f \, c & 0 \end{array} \right) \; \tilde U_e \label{eq:masskneq0} 
\end{equation}
where 
\newpage
\begin{eqnarray}
\zeta = & ({ S \ V'_{16} \over  \phi \ V_{16} }) &  \label{eq:4nu1} \\
b = &  \omega \ \zeta  + 2 \ \sigma \ r \ \epsilon & \nonumber \\
\bar \epsilon^2 = & (\epsilon^\prime)^2 + 9 \, (\kappa_1 - \kappa_2^2) \, \epsilon^2 &  \nonumber \\
C_{2 3} = & \left( 1 + \frac{3}{2} \, \frac{\kappa_1 \, \epsilon^3 \, r^2 \, \sigma \, (3 - 4 \sigma)}{\bar \epsilon^2}\right) - \frac{3 \, (\kappa_1 - \kappa_2^2) \, \epsilon^2 - \kappa_2 \, \epsilon \, \epsilon^\prime}{\bar \epsilon^2} \, r \, \sigma \, \zeta &  \nonumber \\
C_{3 3} = & - \frac{6 \, \kappa_1 \, \epsilon^2 \, r \, \sigma}{\bar \epsilon^2 \, \omega} \, \left( 1 + \frac{3 \, \kappa_1 \, \epsilon^3 \, r^2 \, \sigma \, (1 - \sigma)}{\bar \epsilon^2}\right) + \frac{\kappa_1 \, \epsilon^2 \, r^2 \, \sigma^2}{\bar \epsilon^2 \, \omega} \, \zeta & \nonumber \\
u = &  r \, \sigma \, \epsilon &  \nonumber \\
c = &  \frac{\mu_3 \, V_{16}}{\omega \, m_t \, \phi} &  \nonumber \\
f = &   1 + \frac{3 \epsilon^3 \kappa_1 r^2 \sigma (1 - \sigma)}{\bar \epsilon^2}& \nonumber
\end{eqnarray}
and
\begin{eqnarray}  m' &=& \frac{\lambda^2 v^2 \sin^2\beta~ \omega \phi}
{2 V_{16} V'_{16}} \approx {m_t^2 \ \omega \ \phi  \over
V_{16} \ V'_{16}}  \label{eq:4nu2} 
 \end{eqnarray}
where in the approximation for $m'$ we use
\begin{eqnarray}   m_t (\equiv m_{top}) \approx \lambda \ {v \over \sqrt{2}}
\ \sin\beta ,
\end{eqnarray}  valid at the weak scale.

In addition, for $N_\nu=5$ the off-diagonal piece
   of the mass matrix in eq.(\ref{eq:generalmnu})
   reads
\beq
 - {\tilde \mu}^T \, V^{-1}\: m^T = -m'\; d\; \left(
      \begin{array}{ccc}
                         1 & 0 & (u-r\epsilon/2)\,g \\ 
                         0 & 1 & (u-r\epsilon/2)\,h 
      \end{array} \right),
\label{eq:mass5offdg}
\eeq
with 
\bea
    d&=& \frac{\mu^{\prime}\;V_{16}^\prime}{m_t\;\phi} \\
    g&=& (3\epsilon\kappa_2 + \epsilon^{\prime})/\bar\epsilon^2 \\
    h&=& -3\epsilon\kappa_1/\bar\epsilon^2. 
\eea

\subsection{Three Neutrinos}

Before we discuss the case with non-zero $\kappa_{(1,2)}$, let's recall the problem when $\kappa_{(1,2)} = 0$.
For three active neutrinos with minimal family breaking vevs,  $\langle \phi^2 \rangle, \; \langle S^{2 2} \rangle,\; \langle A^{1 2} \rangle \neq 0$ and $\kappa_1 = \kappa_2 = 0$,  we find (at $M_G$) in
the ($\nu_e,\; \nu_\mu,\; \nu_\tau$) basis
\begin{eqnarray}
m_{\nu} =  m' \; U_e^T \; \left(\begin{array}{ccc}  0 & 0 & 0 \\
                                       0 & b & 1  \\
                                       0 & 1 & 0  \end{array} \right) \; U_e
\end{eqnarray}

$m_{\nu}$ is given in terms of two independent parameters \{ $m', \; b$ \}
(see equations ~\ref{eq:4nu1}, ~\ref{eq:4nu2}).
Note, this theory in principle solves two problems associated with neutrino
masses.   It naturally has small mixing between $\nu_e - \nu_{\mu}$ since
 the mixing angle comes purely from diagonalizing the charged
lepton mass matrix which, like quarks, has small mixing angles.   While, for
$b \leq  1$, $\nu_{\mu} - \nu_{\tau}$ mixing is large without fine tuning.
Also note, in this theory one neutrino (predominantly $\nu_e$) is
massless.

We have checked that in this theory it is possible to simultaneously fit both atmospheric and LSND data.  We however cannot simultaneously fit both solar and atmospheric neutrino data.   As discussed in paper I~\cite{brt} this problem can be solved at the expense of adding a new family symmetry breaking vev~\footnote{This additional vev was necessary in the analysis of Carone and Hall.~\cite{u2neutrino}}
 \begin{eqnarray}  \langle \phi^1 \rangle = \kappa \langle \phi^2\rangle  .
  \label{eq:kappa} \end{eqnarray}

In this paper we consider the most general family symmetry breaking vevs, given in equation ~\ref{eq:vevs}, introducing two new complex parameters  $\kappa_1, \; \kappa_2$.  This allows us to obtain a small mass difference between the first and second mass eigenvalues which was unattainable before in the large mixing limit for $\nu_\mu - \nu_\tau$.  Hence good fits to both solar and atmospheric neutrino data can now be found.  In addition, in the previous section we showed that small values of $\kappa_{1, 2}$ are consistent with good fits for charged fermion masses and mixing angles.  In the next section we discuss these new solutions.   

\section{Neutrino oscillations   [ 3 active only ]}

In this section we consider the solutions to atmospheric and solar neutrino oscillations with three neutrinos.  The mass matrix is given in equation ~\ref{eq:masskneq0} with the parameter $c = 0$.    There are three possible solutions to the solar neutrino data defined as small mixing angle [SMA] MSW, large mixing angle [LMA] MSW or ``Just so'' vacuum oscillations~\cite{massmatrices}.  In all three cases atmospheric neutrino data are predominantly described by $\nu_\mu \rightarrow \nu_\tau$ oscillations.

Instead of fitting the data directly, we compare our models with existing 2 neutrino oscillation fits to the data~\cite{massmatrices}.  We use the latest two neutrino fits to the most recent Super-Kamiokande data for atmospheric neutrino oscillations and the best fits to solar neutrino data including the possibility of ``just so" vacuum oscillations or both large and small angle MSW oscillations~\cite{atmos, solar, massmatrices}.

For atmospheric neutrino oscillations we have evaluated the probabilities
($P(\nu_\mu \rightarrow \nu_\mu)$,   $P(\nu_\mu \rightarrow \nu_x) \ {\rm
with} \ x = \{ e, \ \tau, \ s \}$)  as a function of ${\rm x} \equiv
\rm{Log}[(L/km)/(E/GeV)]$.  In order to smooth out the oscillations we have 
averaged the result over a bin size, $\Delta$x = 0.5.   In  figures 2a and 4a we see that our results are in good agreement with the values of $\delta m^2_{atm}$ and $\sin^2 2 \theta_{atm}$ as given.

For solar neutrinos we plot,  in figures 3(a,b) and 5(a,b), the probabilities  ($P(\nu_e \rightarrow \nu_e)$,  $P(\nu_e \rightarrow \nu_x) \ {\rm with}
\ x = \{ \mu, \ \tau, \ s \}$) for neutrinos produced at the center of the
sun to propagate to the surface (and then without change to earth), as a function of the neutrino energy  E$_\nu$ (MeV).~\footnote{For this calculation use an analytic approximation necessary to account for both large and small oscillation scales. For the details, see the appendix.}  We then compare our model to a 2 neutrino oscillation model with the given parameters.   

\subsection{3 $\nu$ SMA MSW solution}

In tables ~\ref{t:3nm2sma} and ~\ref{t:3nangsma} we give the parameters for the fit corresponding to figures 2(a,b) and 3(a,b).  This model is indistinguishable from the results of the given parameters for 2 neutrino oscillations $\nu_\mu - \nu_\tau$ for atmospheric data and $\nu_e - \nu_{active}$ for solar data.

In order to obtain a SMA MSW solution we need to choose $\kappa_1 = \kappa_2^2$ to high accuracy.  Note this value of $\kappa_{(1,2)}$ corresponds to the only solution obtained previously (in I) with non zero $\kappa$ defined by $\langle \phi^1 \rangle = \kappa \, \langle \phi^2 \rangle$.   In fact, an SU(2) rotation of this case to zero $\langle \phi^1 \rangle$ gives non zero  $\langle S^{1 1} \rangle, \;  \langle S^{1 2}\rangle $ satisfying the relation $\kappa_1 = \kappa_2^2$.  

The parameter $m'$ is determined by the high see-saw scale.  Given $m'$(eqn. ~\ref{eq:4nu2} and table ~\ref{t:3nm2sma}) we find $V_{16} V'_{16}/ \phi = 1.33 \cdot 10^{16}$ GeV which is consistent with the GUT scale.  The large value of $b$ (eqn. ~\ref{eq:4nu1}) results from $S \sim 10 \; \phi$ which is needed in order to have one large and two small eigenvalues.

\protect
\begin{table}
\caption[3]{ 
{\bf Fit to atmospheric and solar neutrino oscillations [3 $\nu$ SMA MSW]}\\\mbox{Initial parameters: ( $\kappa_1 = \kappa_2^2$, $|\kappa_2| = 0.028$   ) }\ \ \ \

$m'$ = 3.35 $\cdot$10$^{-3}$ eV , \ \ $b$ = 15.0,   \ \ $\Phi_b$ =
3.30rad
}
\label{t:3nm2sma}
$$
\begin{array}{|c|c|}
\hline
{\rm Observable} &{\rm Computed \;\; value} \\ 
\hline
\delta m^2_{atm}            &  3.5 \cdot 10^{-3} \ \rm eV^2          \\
\sin^2 2\theta_{atm}            &  0.99        \\
 \delta m^2_{sol}   &  6.3\cdot 10^{-6}  \ \rm eV^2  \\
\sin^2 2\theta_{sol} &  5.2\cdot 10^{-3}         \\
\hline
\end{array}$$
\end{table}

\protect
\begin{table}
\caption[3]{ {\bf Neutrino Masses and Mixings  \ \ [3 $\nu$  SMA MSW]}
\mbox{Mass eigenvalues [eV]: \ \  0.000001, \ 0.0025, \ 0.059 \hspace{1cm}} \\

Magnitude of neutrino mixing matrix  U$_{\alpha \ i}$  \\
\mbox{ $i = 1, \cdots, 3$ -- labels mass eigenstates.}
\mbox{ $\alpha = \{ e, \ \mu, \ \tau \}$ labels flavor eigenstates.}

}
\label{t:3nangsma}
$$
\left[ \begin{array}{ccc}

0.997 &  0.0360 & 0.0599   \\
0.0677     &    0.672     & 0.738      \\
0.0172     &    0.740    & 0.673     \\
\end{array} \right]$$
\end{table}

\renewcommand{\thefigure}{2 \alph{figure}}\setcounter{figure}{0}
\begin{figure}
	\centerline{ \psfig{file=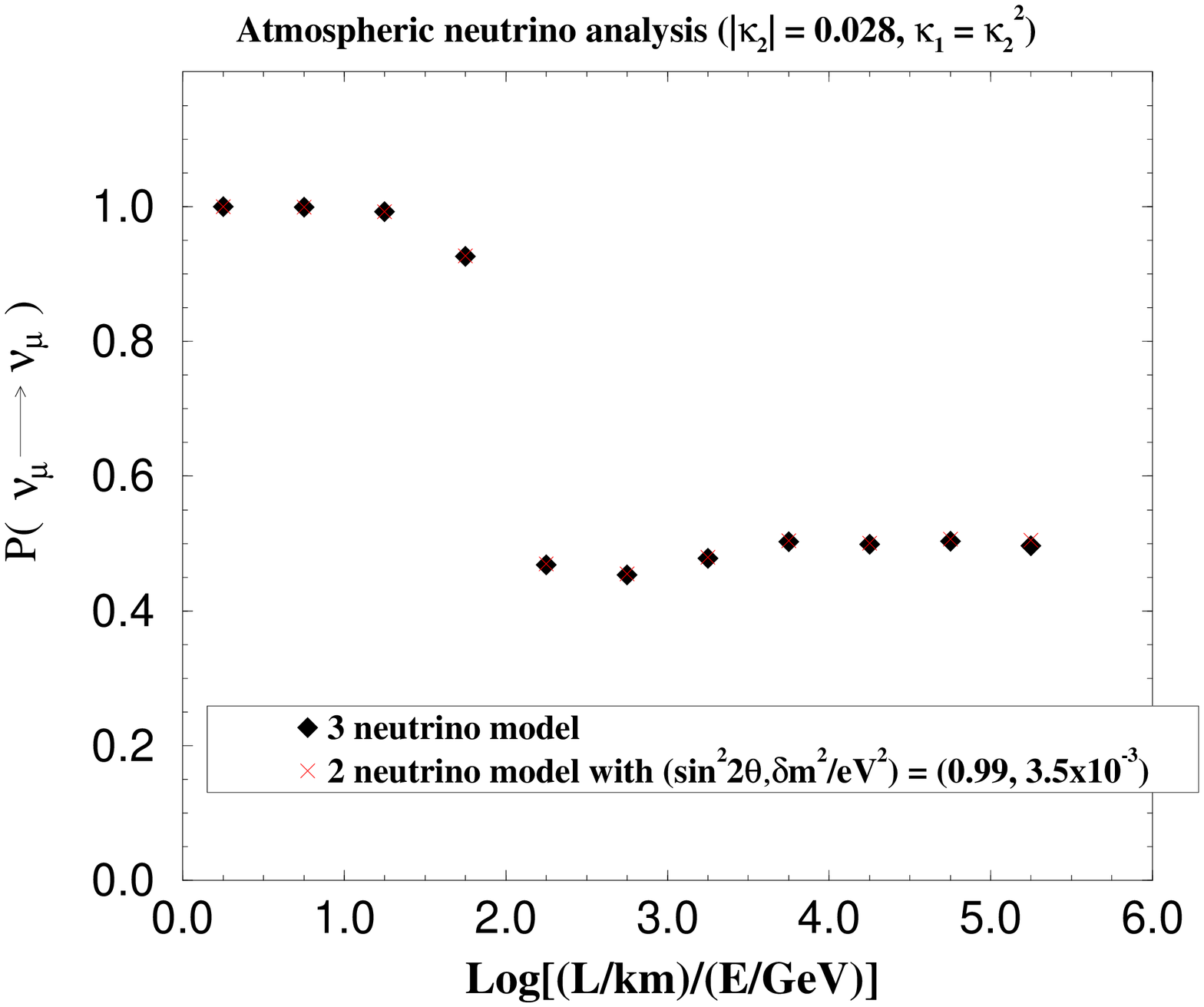,width=9cm,angle=-90}}
\caption{Probability $P(\nu_\mu \longrightarrow \nu_\mu)$ for atmospheric
neutrinos [3 $\nu$  SMA MSW]. For this analysis, we neglect the matter effects.}
\end{figure}
\begin{figure}
	\centerline{ \psfig{file=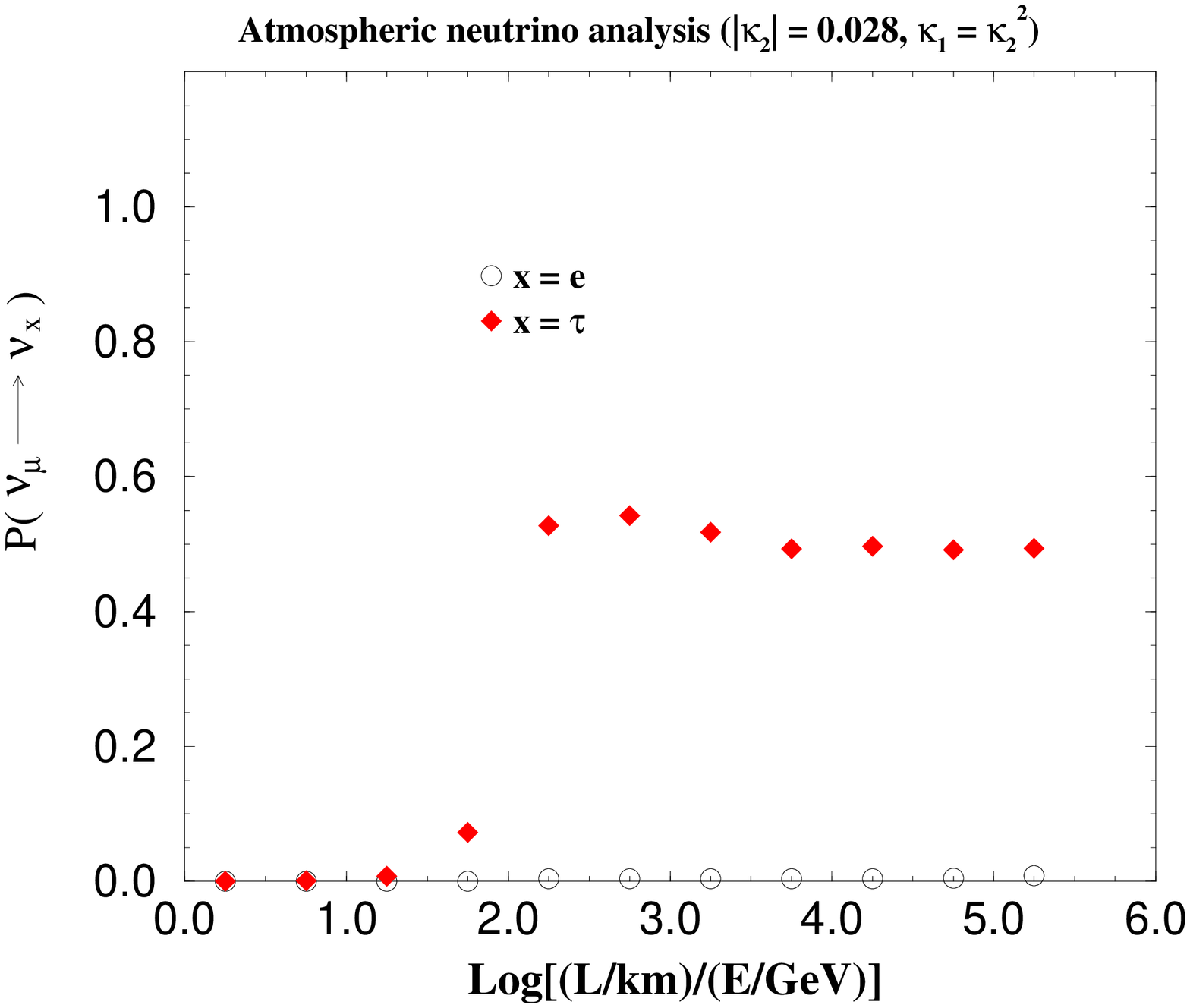,width=9cm,angle=-90}}
\caption{Probabilities $P(\nu_\mu \longrightarrow \nu_x)$ ($x=e$, $\tau$ and
$s$) for atmospheric neutrinos [3 $\nu$  SMA MSW].}
\end{figure}

\renewcommand{\thefigure}{3 \alph{figure}}\setcounter{figure}{0}
\begin{figure}
	\centerline{ \psfig{file=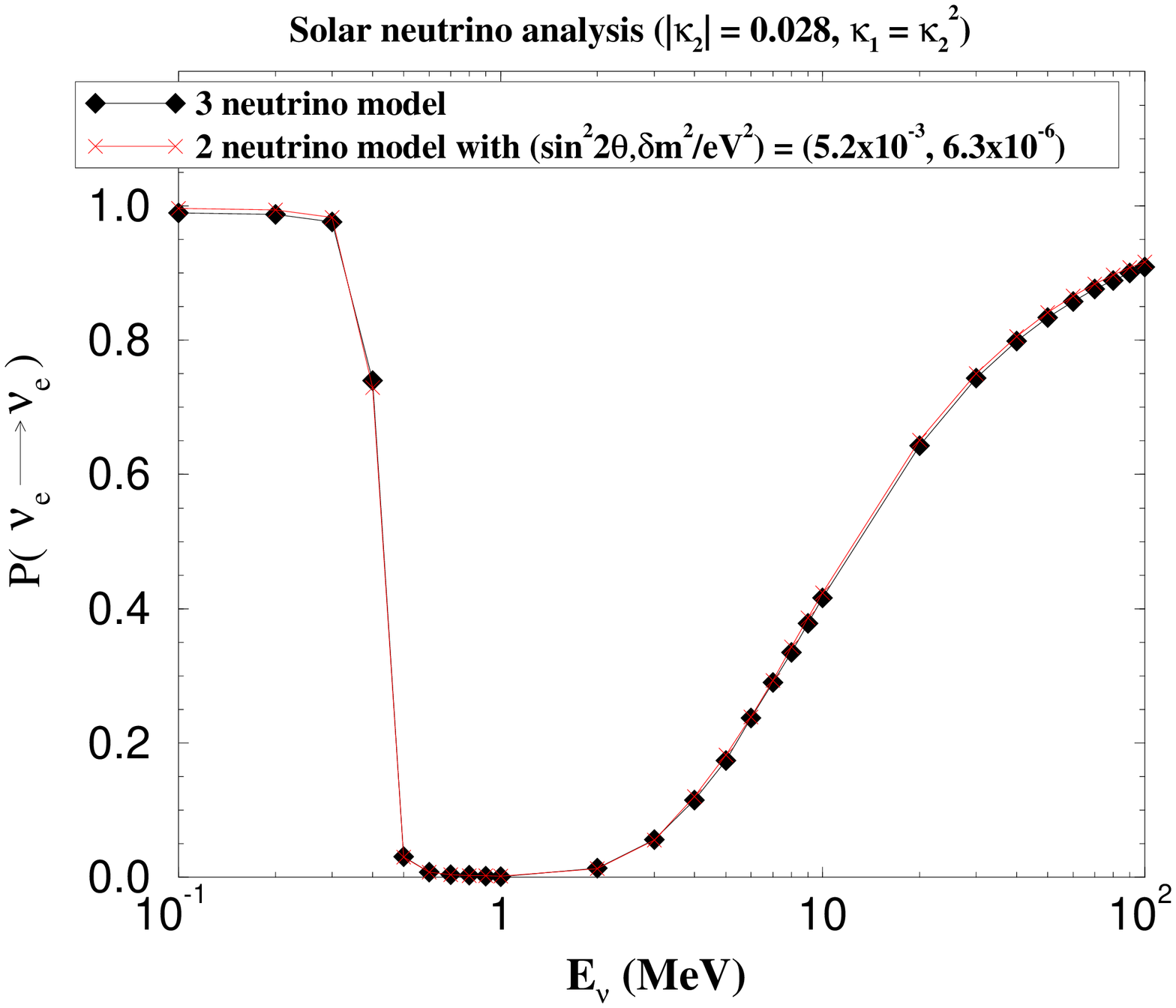,width=9cm,angle=-90}}
\caption{Probability $P(\nu_e \longrightarrow \nu_e)$ for solar neutrinos [3 $\nu$  SMA MSW].}
\end{figure}
\begin{figure}
	\centerline{ \psfig{file=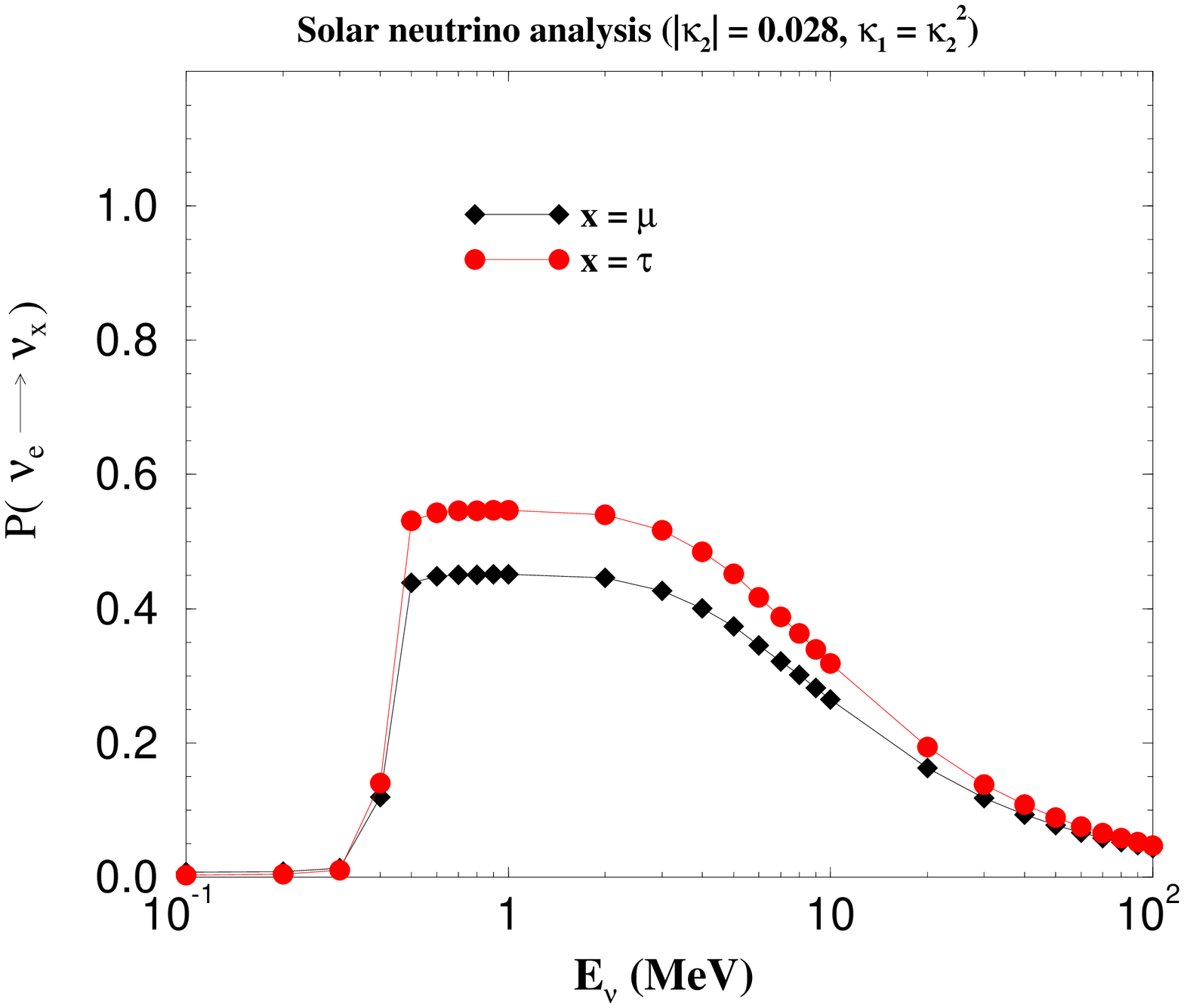,width=9cm,angle=-90}}
\caption{Probabilities $P(\nu_e \longrightarrow \nu_x)$ ($x=\mu$, $\tau$ and
$s$) for solar neutrinos [3 $\nu$  SMA MSW].}
\end{figure}

\subsection{3 $\nu$ LMA MSW solution}

In figures 4(a,b) and 5(a,b) we present the comparison to a two neutrino oscillation model for atmospheric and solar neutrino data (see also tables ~\ref{t:3nm2lma} and ~\ref{t:3nanglma}).  For atmospheric data the fit is good for values of \newline $Log((L/km)/(E/GeV)) \leq 4$ (see figures 4(a,b))
where the oscillations are predominantly given by $\nu_\mu - \nu_\tau$.   For larger $x \geq 4$ the probability $P(\nu_\mu \rightarrow \nu_\mu)$ is significantly smaller ($\sim$ 30 \%) in our model than in a two neutrino model.  This is due to the onset of significant $\nu_\mu \rightarrow \nu_e$ oscillations.  These larger values of $x$ may be accessible in atmospheric oscillations.   The maximum distance $L$ for neutrinos of order 13,000 km, for upward going neutrinos, and the minimum detectable energy of order 0.1 GeV, corresponds to a value of $x_{MAX} \sim 5$.  On the other hand, it would require a much more detailed analysis to determine whether our model is consistent with the data for fully contained events in the sub GeV ( $<$ 1.33 GeV ) regime.  We also note that this effect has been considered, in a recent analysis by Peres and Smirnov~\cite{peressmirnov}, as a possible tool to distinguish the LMA MSW solution from the other solutions to the solar neutrino problem.

\protect
\begin{table}
\caption[3]{ 
{\bf Fit to atmospheric and solar neutrino oscillations [3 $\nu$ LMA MSW]}\\\mbox{Initial parameters: ( $|\kappa_1| = 0.055$, $|\kappa_2| = 0.31$ ) }\ \ \ \

$m'$ = 4.93 $\cdot$10$^{-2}$ eV , \ \ $b$ = 0.84,   \ \ $\Phi_b$ =
-0.41 rad
}
\label{t:3nm2lma}
$$
\begin{array}{|c|c|}
\hline
{\rm Observable} &{\rm Computed \;\; value} \\ 
\hline
\delta m^2_{atm}            &  3.7 \cdot 10^{-3} \ \rm eV^2          \\
\sin^2 2\theta_{atm}            &  0.99        \\
 \delta m^2_{sol}   &  2.3\cdot 10^{-5}  \ \rm eV^2  \\
\sin^2 2\theta_{sol} &  0.77        \\
\hline
\end{array}$$
\end{table}

\protect
\begin{table}
\caption[3]{ {\bf Neutrino Masses and Mixings  \ \ [3 $\nu$ LMA MSW]}
\mbox{Mass eigenvalues [eV]: \ \  0.002, \ 0.005, \ 0.061 \hspace{1cm}} \\

Magnitude of neutrino mixing matrix  U$_{\alpha \ i}$  \\
\mbox{ $i = 1, \cdots, 3$ -- labels mass eigenstates.}
\mbox{ $\alpha = \{ e, \ \mu, \ \tau \}$ labels flavor eigenstates.}

}
\label{t:3nanglma}
$$
\left[ \begin{array}{ccc}

0.857 &  0.513 & 0.049   \\
0.368     &    0.563     & 0.740      \\
0.362     &    0.648    & 0.671     \\
\end{array} \right]$$
\end{table}

\renewcommand{\thefigure}{4 \alph{figure}}\setcounter{figure}{0}
\begin{figure}
	\centerline{ \psfig{file=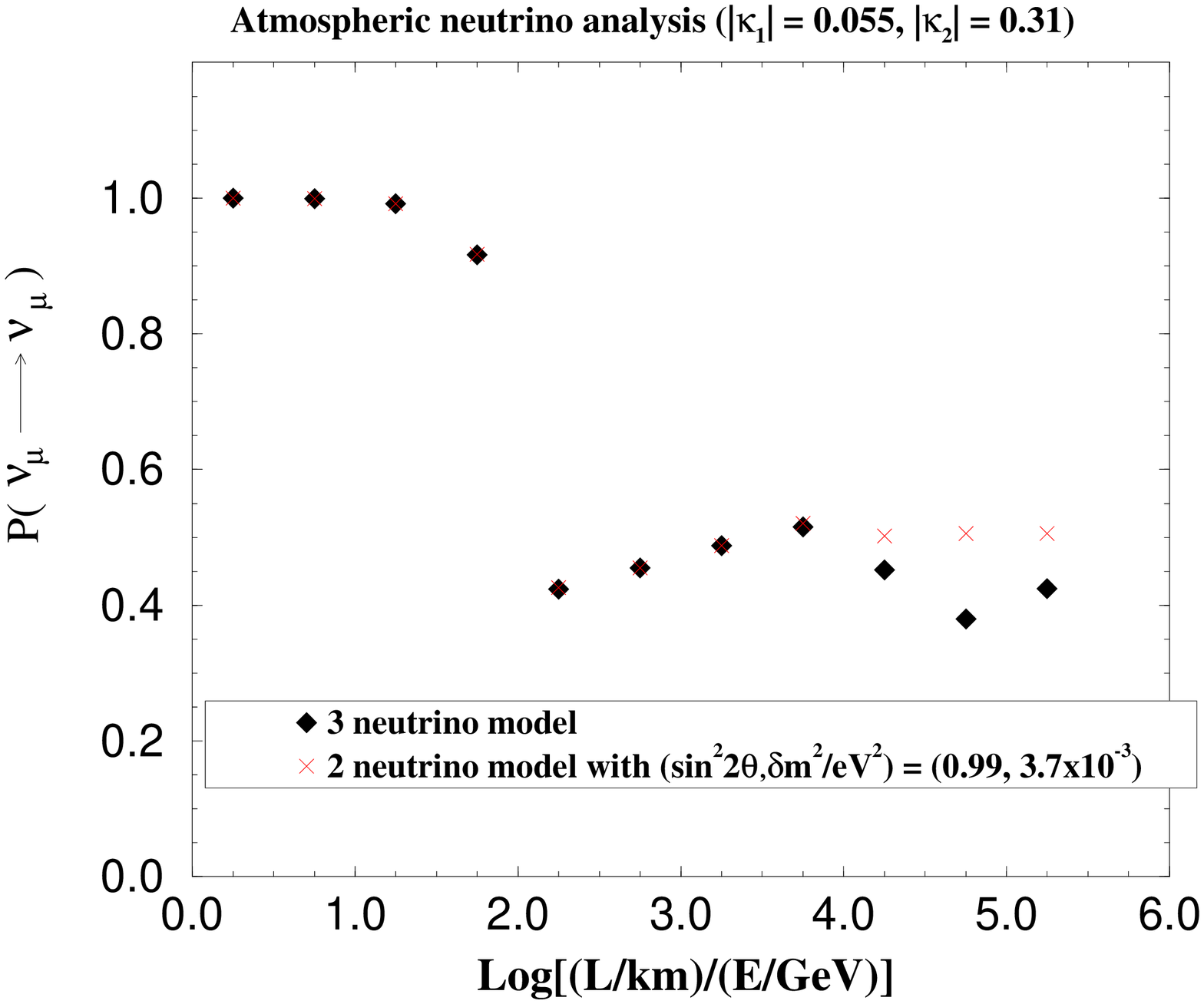,width=9cm,angle=-90}}
\caption{Probability $P(\nu_\mu \longrightarrow \nu_\mu)$ for atmospheric
neutrinos [3 $\nu$ LMA MSW]. For this analysis, we neglect the matter effects.}
\end{figure}
\begin{figure}
	\centerline{ \psfig{file=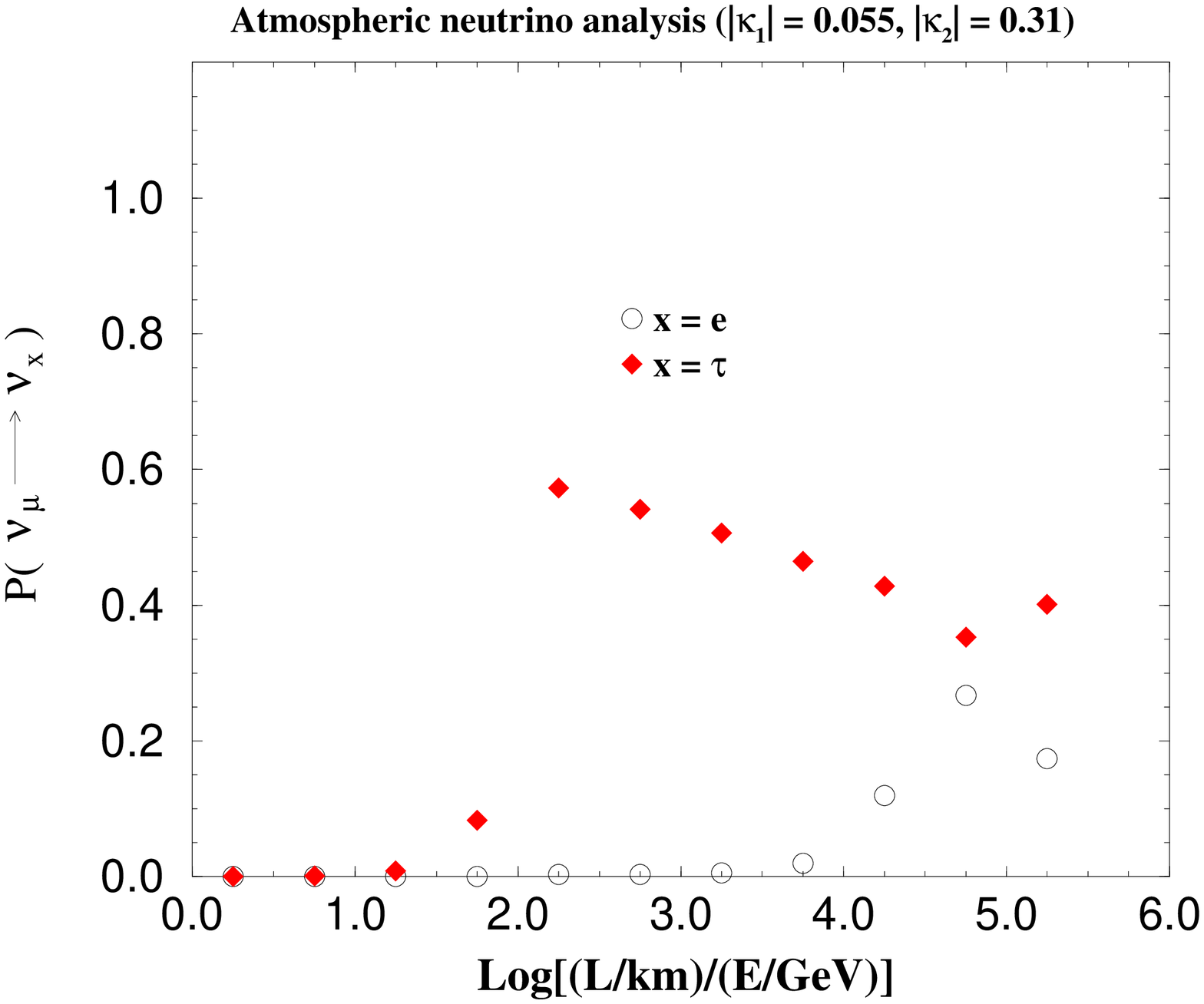,width=9cm,angle=-90}}
\caption{Probabilities $P(\nu_\mu \longrightarrow \nu_x)$ ($x=e$, $\tau$ and
$s$) for atmospheric neutrinos [3 $\nu$ LMA MSW].}
\end{figure}

\renewcommand{\thefigure}{5 \alph{figure}}\setcounter{figure}{0}
\begin{figure}
	\centerline{ \psfig{file=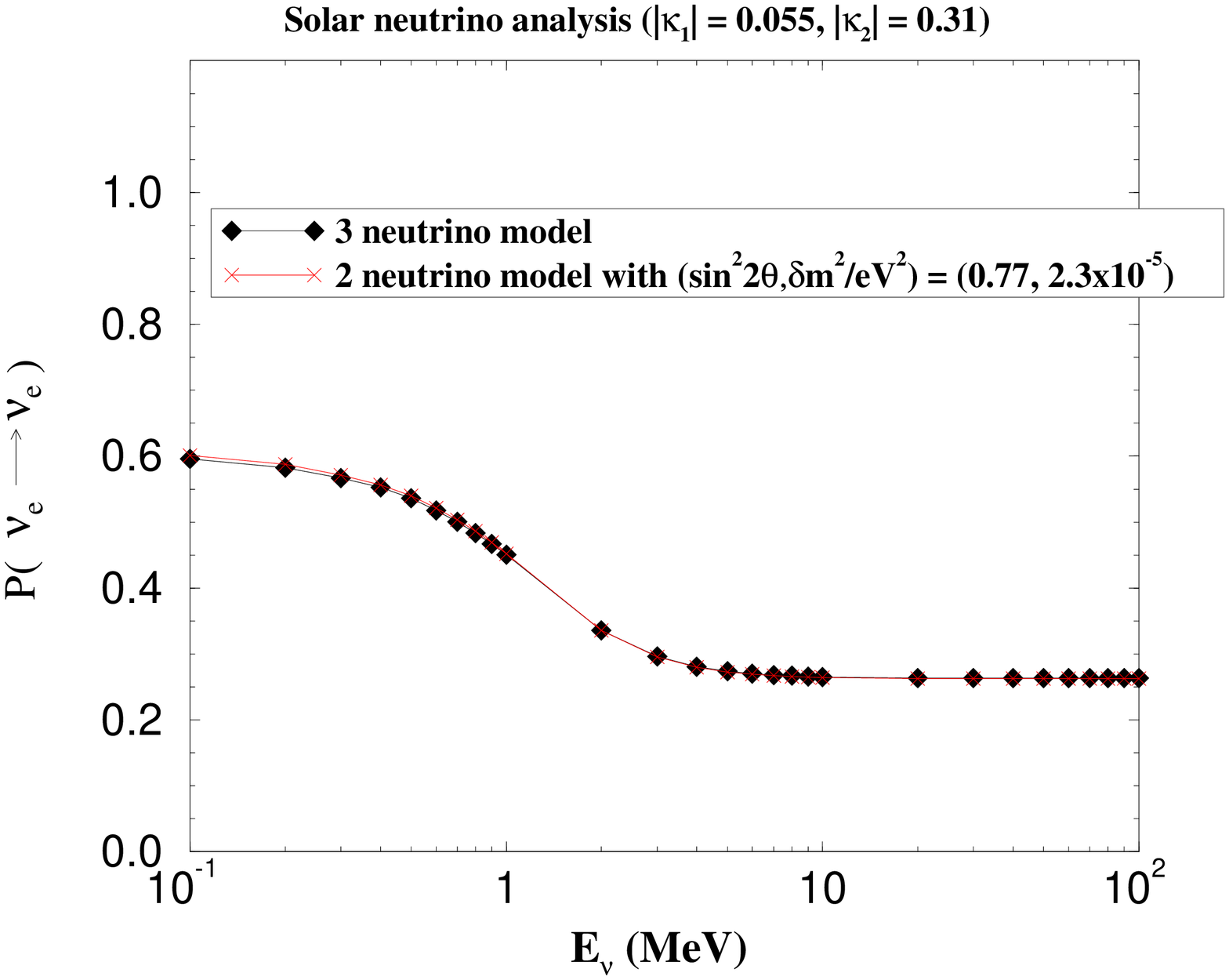,width=9cm,angle=-90}}
\caption{Probability $P(\nu_e \longrightarrow \nu_e)$ for solar neutrinos [3 $\nu$ LMA MSW].}
\end{figure}
\begin{figure}
	\centerline{ \psfig{file=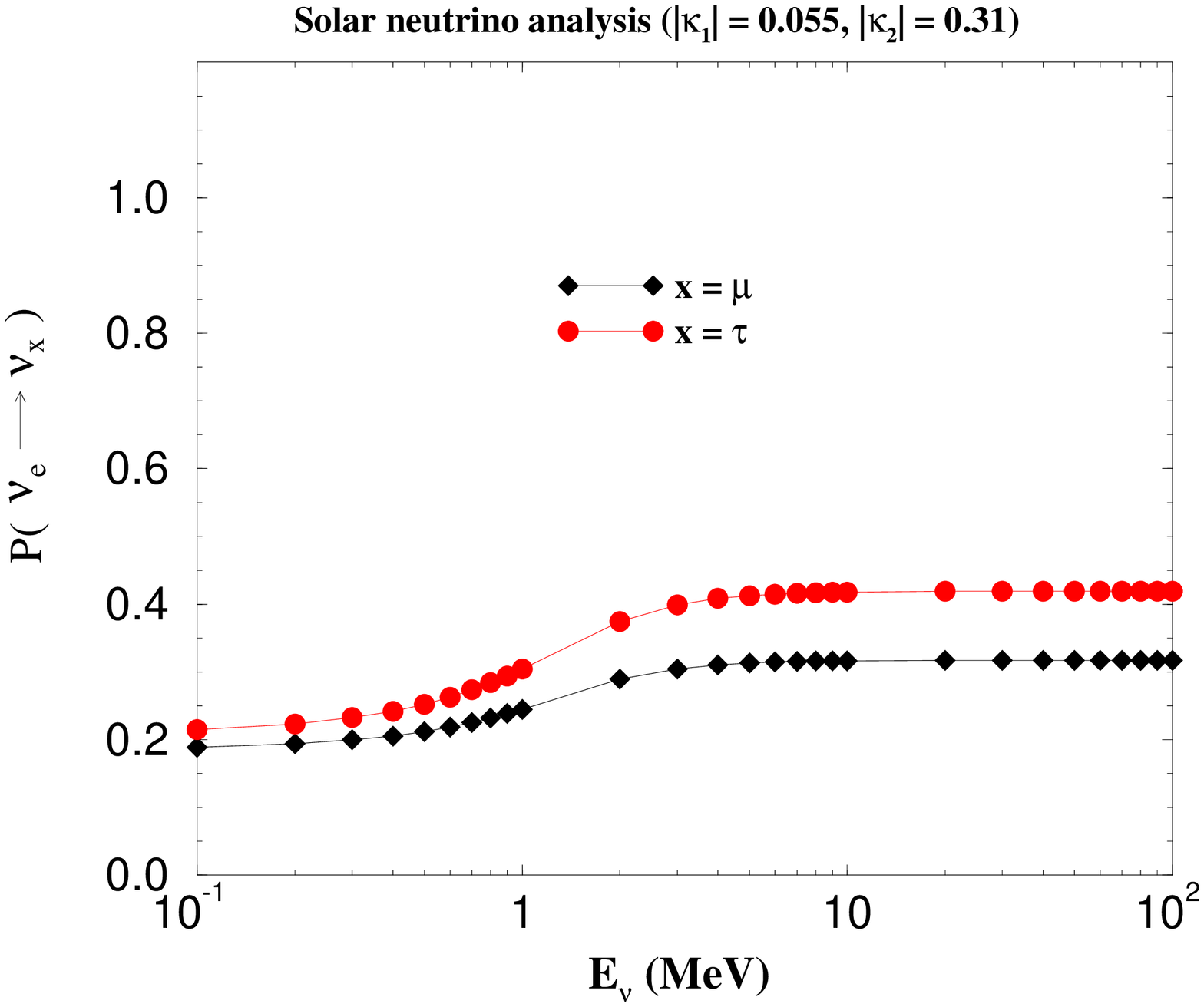,width=9cm,angle=-90}}
\caption{Probabilities $P(\nu_e \longrightarrow \nu_x)$ ($x=\mu$, $\tau$ and
$s$) for solar neutrinos [3 $\nu$ LMA MSW].}
\end{figure}

A large mixing angle oscillation solution is obtained by tuning the lightest two neutrinos to be approximately degenerate with a near bi-maximal mixing matrix (see tables ~\ref{t:3nm2lma} and ~\ref{t:3nanglma}), where the bi-maximal mixing matrix is given by~\cite{bargeretal} 
\bea   |U_{\alpha \ i}| = &
\left[ \begin{array}{ccc}
0.71 &  0.71 & 0.0   \\
0.5     &    0.5     & 0.71      \\
0.5     &    0.5    & 0.71   \\
\end{array} \right]   &
\eea
Note, a major difference in our case is the non-zero value for $U_{\nu_e 3} \sim .049$.  However the constraint $U_{\nu_e 3} = 0$ chosen to satisfy CHOOZ~\cite{chooz} is much too strong.   It is easy to see that our model is consistent with the null results of CHOOZ, i.e.
\bea  
P(\nu_e \rightarrow \nu_e) =& 1 - 4 |U_{\nu_e 3}|^2 (1 - |U_{\nu_e 3}|^2) sin^2 (\frac{1.27 \delta m_{atm}^2(eV^2) L(km)}{E_\nu (GeV)}) &  > 0.98  \nonumber \\
& &  
\eea
for values of $|U_{\nu_e 3}| \le 0.16$~\cite{bargerwhisnant}.
Finally the parameter $b \sim 1$ requires no fine tuning and given $m'$ we find the high see-saw scale $V_{16} V'_{16}/ \phi = 8.78 \cdot 10^{14}$ GeV.

\subsection{3 $\nu$ ``Just So" Vacuum solution}
A vacuum solution is obtained by tuning the lightest two neutrinos to be even more degenerate than in the previous LMA MSW case.  Our results are given in tables ~\ref{t:3nm2vac} and ~\ref{t:3nangvac}.  We have not given any figures since the results are standard vacuum oscillations.  Once again we obtain a near
bi-maximal mixing matrix~\cite{bargeretal}  with however  $U_{\nu_e 3} \sim .049$.  Nevertheless this model is consistent with CHOOZ data ~\cite{chooz} (see the discussion of this in the LMA MSW case).
Finally given the overall scale $m'$ we determine the high energy scale to be
$V_{16} V'_{16}/ \phi = 2.27 \cdot 10^{15}$ GeV  and $b \sim 1$.

\protect
\begin{table}
\caption[3]{ 
{\bf Fit to atmospheric and solar neutrino oscillations [3 $\nu$ Vacuum]}\\\mbox{Initial parameters: ( $|\kappa_1| = 0.004$, $|\kappa_2| = 0.025$ ) }\ \ \ \

$m'$ = 2.92 $\cdot$10$^{-2}$ eV , \ \ $b$ = 1.73,   \ \ $\Phi_b$ =
-0.33 rad
}
\label{t:3nm2vac}
$$
\begin{array}{|c|c|}
\hline
{\rm Observable} &{\rm Computed \;\; value} \\ 
\hline
\delta m^2_{atm}            &  3.5 \cdot 10^{-3} \ \rm eV^2          \\
\sin^2 2\theta_{atm}            &  0.99        \\
 \delta m^2_{sol}   &  7.9\cdot 10^{-11}  \ \rm eV^2  \\
\sin^2 2\theta_{sol} &  0.97        \\
\hline
\end{array}$$
\end{table}

\protect
\begin{table}
\caption[3]{ {\bf Neutrino Masses and Mixings  \ \ [3 $\nu$  Vacuum]}
\mbox{Mass eigenvalues [eV]: \ \  0.00106037, \ \  0.00106041  , \ 0.059 \hspace{1cm}} \\

Magnitude of neutrino mixing matrix  U$_{\alpha \ i}$  \\
\mbox{ $i = 1, \cdots, 3$ -- labels mass eigenstates.}
\mbox{ $\alpha = \{ e, \ \mu, \ \tau \}$ labels flavor eigenstates.}

}
\label{t:3nangvac}
$$
\left[ \begin{array}{ccc}

0.759 &  0.649 & 0.049   \\
0.429     &    0.513     & 0.744      \\
0.489     &    0.563    & 0.667     \\
\end{array} \right]$$
\end{table}

In the next section we discuss a four neutrino solution to atmospheric, solar and LSND neutrino data in the theory with $\kappa_{(1,2)} \neq 0$.

 \section{Neutrino oscillations [3 active + 1 sterile]}

In the four neutrino case the mass matrix (at $M_G$) is given
by equation ~\ref{eq:masskneq0}  with $c \neq 0$.

As in the previous case of three neutrinos, we compare our model with two-neutrino oscillation models which have already been fit to the data ~\cite{solar, atmos, massmatrices}.     The results for our best fit are found in tables ~\ref{t:4numass2} and ~\ref{t:4nuangles} and figures 6(a,b), 7(a,b) and 8.

In  fig. 6a  we evaluate $P(\nu_\mu \rightarrow \nu_\mu)$ where we also include a multiplicative fudge factor $\alpha$.    This is justified by the theoretical uncertainty in the normalization of the incident $\nu_\mu$ flux.  Recall the observed number of muon neutrinos is given by
\bea
N(\nu_\mu) =  & N_0(\nu_\mu) \;  P(\nu_\mu \rightarrow \nu_\mu)   &
\eea
where $ N_0(\nu_\mu) $ is the theoretically expected incident neutrino flux which has an uncertainty of order 20\%.  We let $N_0(\nu_\mu) =  N_{fit} \cdot \alpha$ where $N_{fit}$ is the value used for the neutrino flux when fitting the data.

We see that our result is in good agreement with the
values of $\delta m^2_{atm} = 3.5 \cdot 10^{-3}$eV$^2$ and $\sin^2 2 \theta_{atm} = 1.0$ with $\alpha = 1.04$.

In fig. 6b we see that the atmospheric neutrino deficit 
is predominantly due to the maximal mixing between $\nu_\mu - \nu_\tau$, as in
the case with $\kappa_1 = \kappa_2 = 0$.  However, in the case with $\kappa_1 = \kappa_2 = 0$ there is also a significant ($\sim$ 10\%
effect)  oscillation of $\nu_\mu - \nu_s$.  In this case, that effect has vanished.   This also means that sterile neutrinos will not come into thermal equilibrium in the early universe, due to the small mixing angle.  Hence, at the  nucleosynthesis epoch this model has only three neutrino species in thermal equilibrium.

 \protect
\begin{table}
\caption[3]{
{\bf Fit to atmospheric, solar  and LSND neutrino oscillations 
[4 neutrinos \  SMA MSW + LSND]} \\\
\mbox{Initial parameters: $|\kappa_1| = 0.0001, \; |\kappa_2| = 0.002$ }\ \  \ \hspace{1in}

$m' = 0.979$ eV , \ $b$ = -0.054, \ $c$ = 0.101,\ $\Phi_b$ = 5.59rad
}
\label{t:4numass2}
$$
\begin{array}{|c|c|}
\hline
{\rm Observable} &{\rm Computed \;\; value} \\
\hline
\delta m^2_{atm}            &  3.5 \cdot 10^{-3} \ \rm eV^2          \\
\sin^2 2\theta_{atm}            &  1.0        \\
 \delta m^2_{sol}   &  5.0\cdot 10^{-6}  \ \rm eV^2  \\
\sin^2 2\theta_{sol} &  3.0 \cdot 10^{-3}         \\
\delta m^2_{LSND} &  0.53  \\
\sin^2 2\theta_{LSND} &  0.018  \\
\hline
\end{array}$$
\end{table}

\protect
\begin{table}
\caption[3]{
{\bf Neutrino Masses and Mixings [4 neutrinos \  SMA MSW + LSND]} \\

\mbox{Mass eigenvalues [eV]: \ \  0.00002, \ 0.0022, \ 0.7248, \ 0.7272 \hspace{1cm}} \\
\mbox{Magnitude of neutrino mixing matrix  U$_{\alpha i}$ \hspace{1.7cm}}\\
\mbox{ $i = 1, \cdots, 4$ -- labels mass eigenstates. \hspace{1.5cm}} \\
\mbox{ $\alpha = \{ e, \ \mu, \ \tau, \ s \}$ labels flavor eigenstates.}
}
\label{t:4nuangles}
$$
\left[ \begin{array}{cccc}
0.997       &  0.0254      & 0.0480      & 0.0482     \\
0.0703      &    0.1079     & 0.7022       & 0.7003      \\
0.273\cdot 10^{-3}   & 0.0292     & 0.7053    & 0.7083     \\
0.0181  & 0.9934    & 0.0852   &  0.0745 \\
\end{array} \right]$$
\end{table}

\renewcommand{\thefigure}{6 \alph{figure}}\setcounter{figure}{0}
\begin{figure}
	\centerline{ \psfig{file=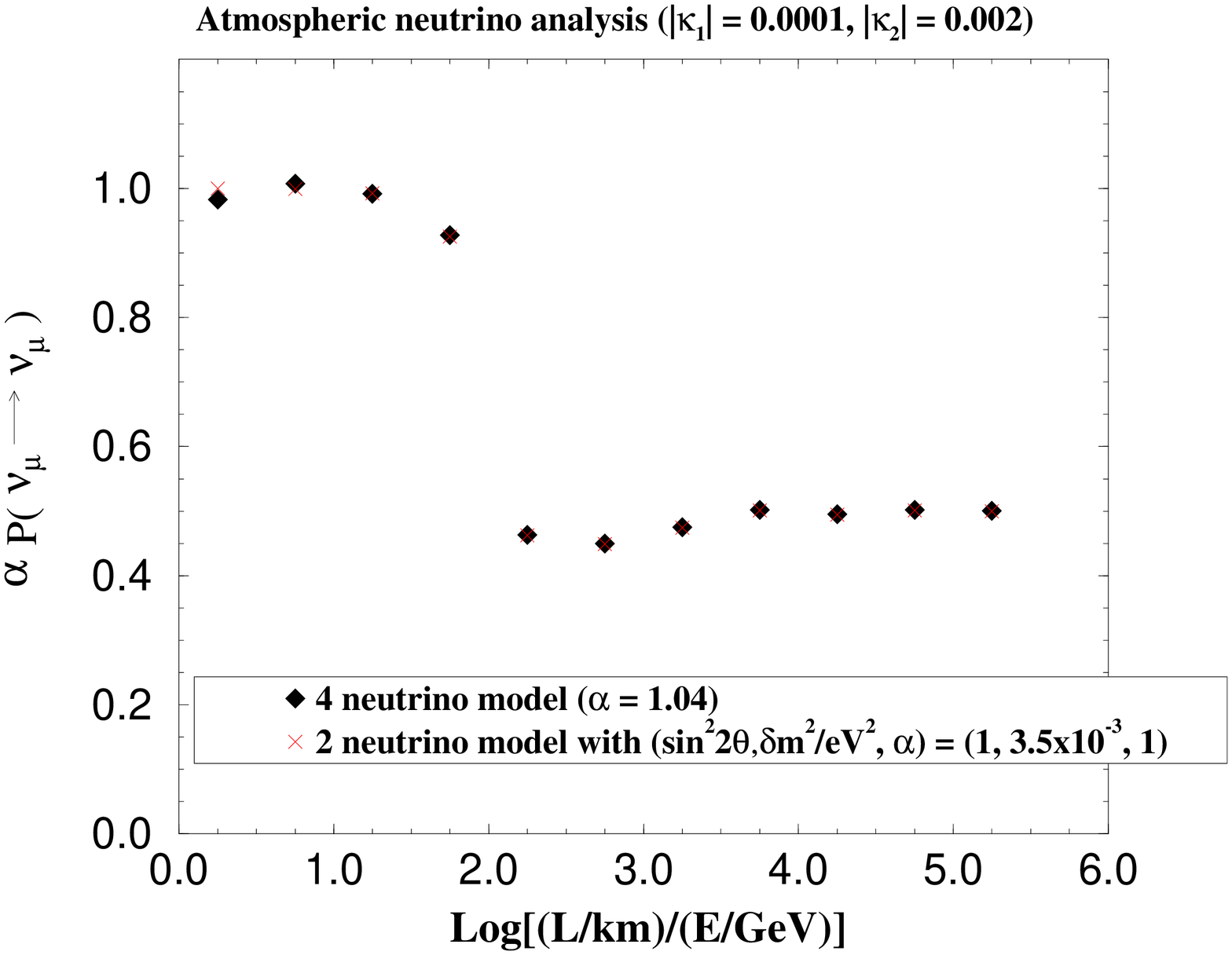,width=9cm,angle=-90}}
\caption{Probability $P(\nu_\mu \longrightarrow \nu_\mu)$ for atmospheric neutrinos multiplied by $\alpha$, a fudge factor introduced to account for the uncertainty in the normalization of the incident $\nu_\mu$ flux.  For this analysis, we neglect matter effects.}
\end{figure}
\begin{figure}
	\centerline{ \psfig{file=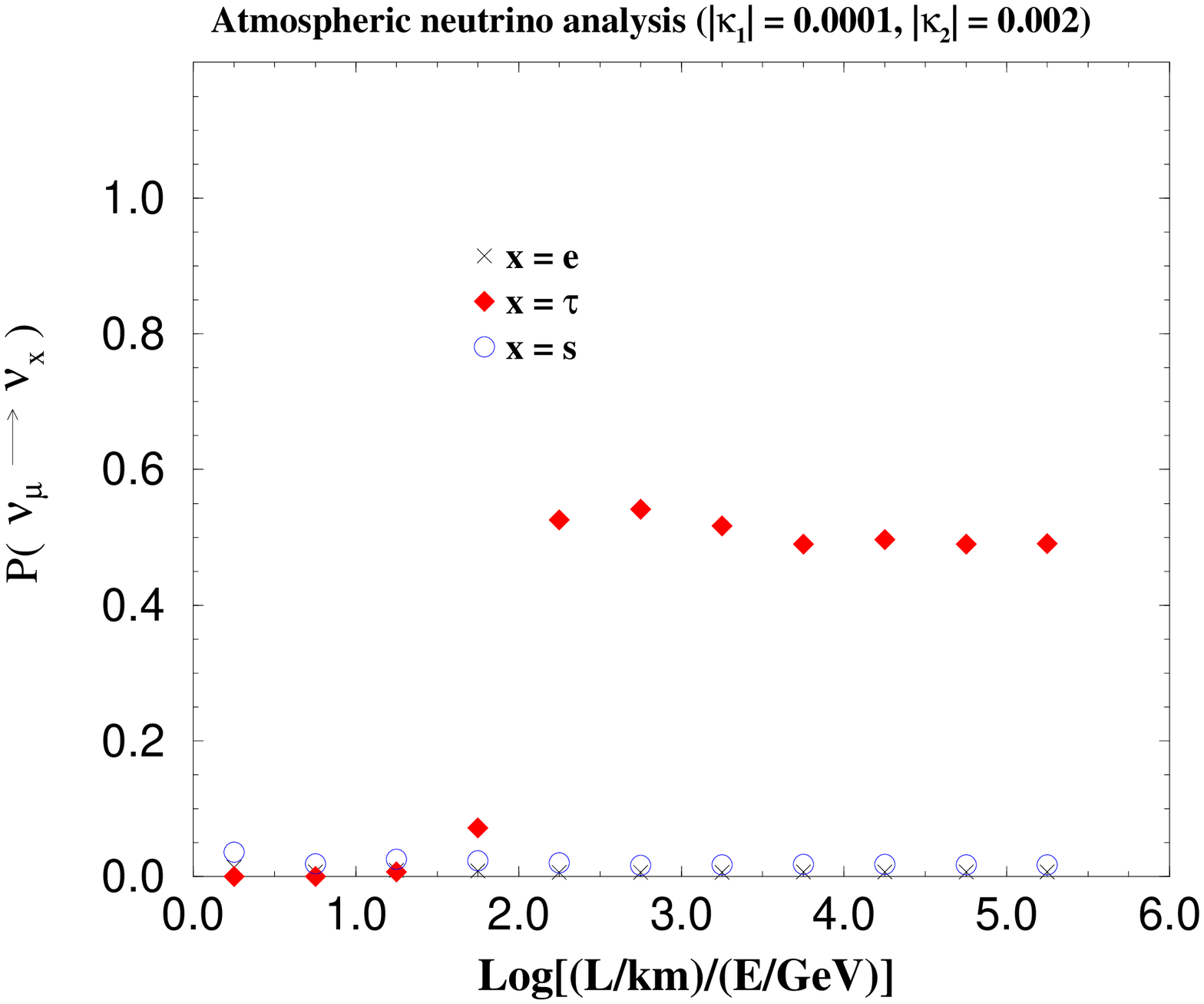,width=9cm,angle=-90}}
\caption{Probabilities $P(\nu_\mu \longrightarrow \nu_x)$ ($x=e$, $\tau$ and
$s$) for atmospheric neutrinos}
\end{figure}

For solar neutrinos we see in fig. 7a  that our model reproduces the neutrino results for $\delta m^2_{sol} = \delta m^2_{12} = 5\times 10^{-6}$ eV$^2$ and
a 2 neutrino mixing angle $\sin^2 2 \theta_{sol} =  3 \times 10^{-3}$.
The solar neutrino deficit is predominantly due to the small mixing angle MSW solution for $\nu_e - \nu_s$ oscillations.  The results are summarized in tables
~\ref{t:4numass2} and ~\ref{t:4nuangles}.

A naive definition of the effective solar mixing angle is given by
\begin{eqnarray} \sin^2 2 \theta_{12} \equiv  4 \ \| U_{e 1} \|^2 \ \|
U_{e 2} \|^2 .
\end{eqnarray}
We note that the naive definition of $\sin^2 2 \theta_{12} $
underestimates the value of the effective 2 neutrino
mixing angle.  The fit value corresponds to $\sin^2 2 \theta_{12} = 2.6 \times 10^{-3}$.

In fig 7b we see that oscillations into any active neutrino is substantially suppressed.  This is unlike the case with $\kappa_1 = \kappa_2 = 0$   where there is also a significant ($\sim$ 8\%) probability for  $\nu_e \rightarrow \nu_\mu$.  

\renewcommand{\thefigure}{7 \alph{figure}}\setcounter{figure}{0}
\begin{figure}
	\centerline{ \psfig{file=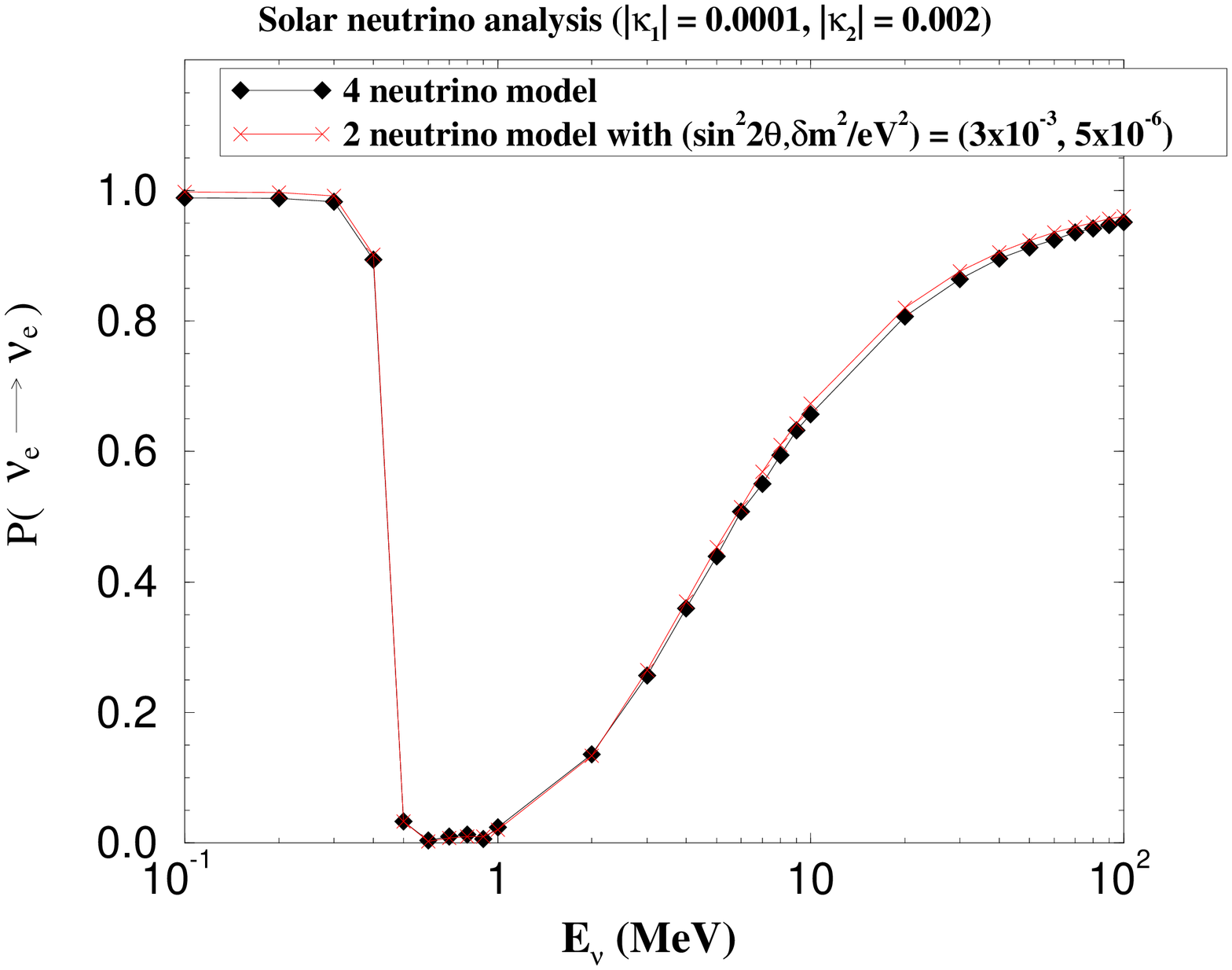,width=9cm,angle=-90}}
\caption{Probability $P(\nu_e \longrightarrow \nu_e)$ for solar neutrinos}
\end{figure}
\begin{figure}
	\centerline{ \psfig{file=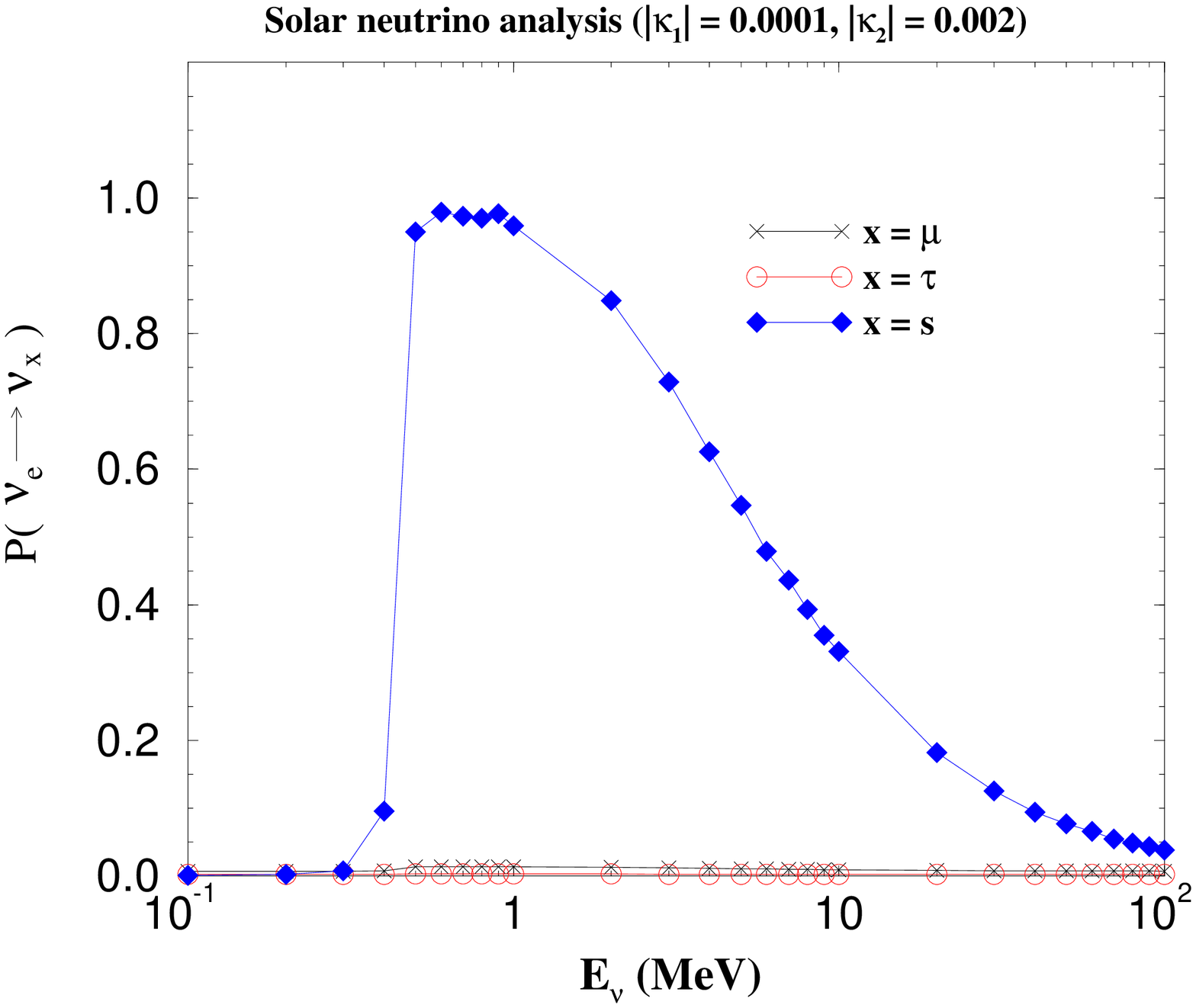,width=9cm,angle=-90}}
\caption{Probabilities $P(\nu_e \longrightarrow \nu_x)$ ($x=\mu$, $\tau$ and
$s$)
for solar neutrinos}
\end{figure}

Finally with non vanishing $\kappa_{1,2}$ we are now able to
simultaneously fit atmospheric, solar and LSND data.  This result is shown in figure 8 where we plot the probability $P(\nu_\mu \longrightarrow \nu_e)$ as a function of neutrino energy relevant for LSND for our model compared to a two neutrino model with $sin^2 2\theta = 0.018$ and $\delta m^2 = 0.53$ eV$^2$ in the LSND allowed region~\cite{LSND}.~\footnote{Note, the probability for ${\bar \nu}_\mu \rightarrow {\bar \nu}_e$ oscillations is almost identical.} This is in contrast to 
the case $\kappa_{1,2} = 0$  (paper I) where this was not possible.  
\renewcommand{\thefigure}{8}\setcounter{figure}{0}
\begin{figure}
	\centerline{ \psfig{file=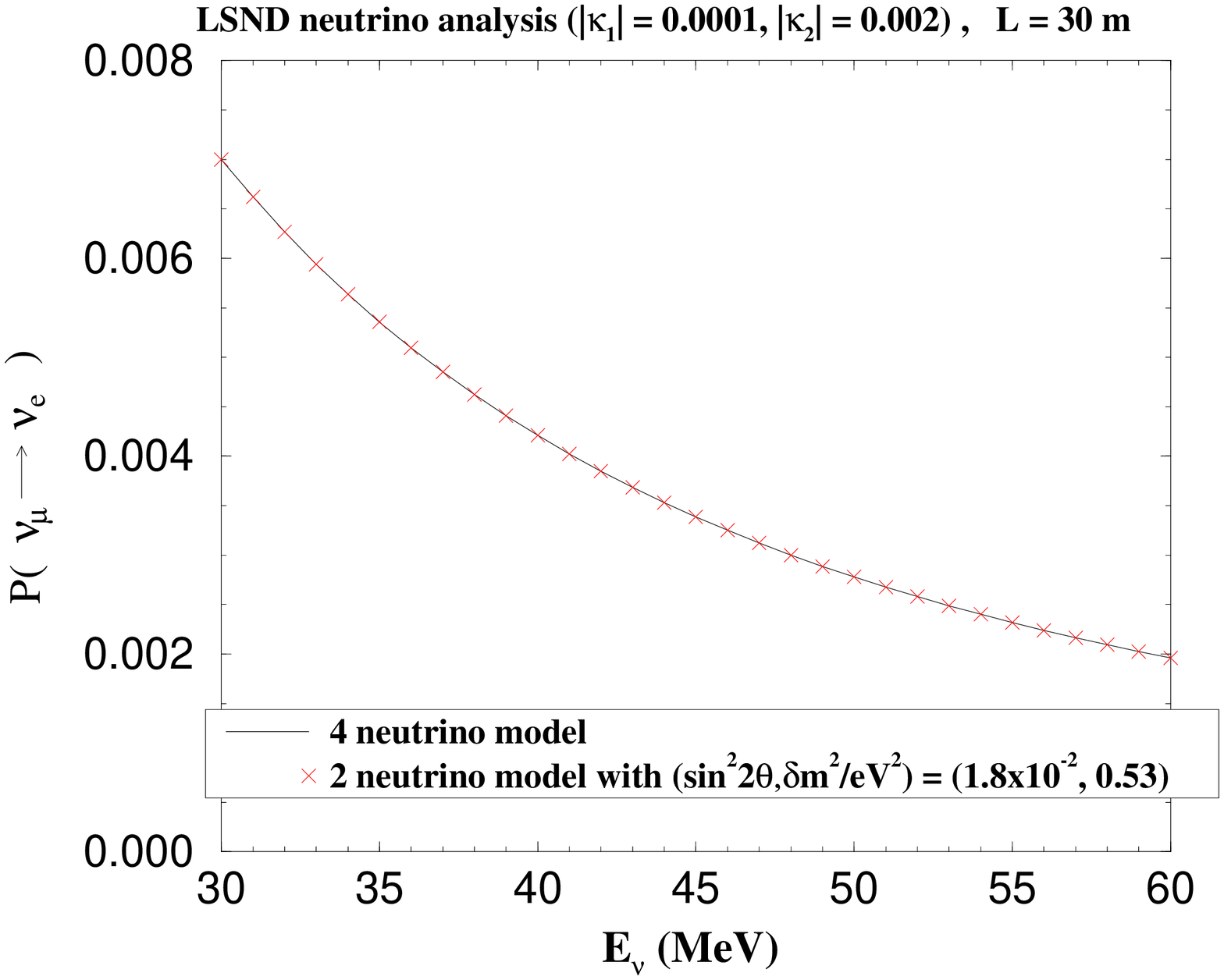,width=9cm,angle=-90}}
\caption{Probability $P(\nu_\mu \longrightarrow \nu_e)$ for LSND energies.}
\end{figure}

We now consider whether the parameters necessary for the fit make sense.
We have three arbitrary parameters.  We have taken $b$ and $c$ complex,
while any phase for $m'$ is unobservable.
A large mixing angle for $\nu_\mu - \nu_\tau$ oscillations is
obtained with $|b| \sim 0.05$ [table ~\ref{t:4numass2}].  This does not require any fine tuning; it is consistent with ${S \ V'_{16} \over \phi \ V_{16}} \sim 0.17$ which, taking into account Yukawa couplings, is perfectly natural (see eqn. ~\ref{eq:4nu1}).  The parameter $c$ [eqn. ~\ref{eq:masskneq0} and table ~\ref{t:4numass2}] $\approx 0.10 \approx { \mu_3 \, V_{16} \over \omega \,
m_t \ \phi}$ implies $ \mu_3 \sim 26 (\frac{\phi}{V_{16}})$ GeV.   Considering that the standard $\mu$ parameter  (see the parameter list in the captions to table \ref{t:fit4nu}) with value $\mu = 110$ GeV and $\mu_3$ [eqn. ~\ref{eq:mu'}] may have similar origins, both generated once SUSY is spontaneously broken, we feel that it is natural to have a light sterile neutrino.  Lastly consider the overall scale of symmetry breaking, i.e. the see-saw scale.  We have $ m' =  0.979 $ eV [table ~\ref{t:4numass2}] $\approx {m_t^2 \ \omega \ \phi  \over  V_{16} \ V'_{16}}$.  Thus we find $ {V_{16} \ V'_{16} \over \phi} \sim  {m_t^2 \ \omega \over m'} \sim 4.66 \times 10^{13}$ GeV.  This is admittedly somewhat small but perfectly reasonable for $\langle \overline{16} \rangle \sim \langle \phi^2 \rangle \sim M_G$ once the implicit Yukawa couplings are taken into account.

\section{Neutrino oscillations [3 active + 2 sterile]}

In this case we have $\mu^\prime \neq 0, \; \mu_3 = 0$ (see eqn. ~\ref{eq:mu}) where $\mu^\prime$ sets the scale for the terms $(m_\nu)_{4 \alpha}, \; (m_\nu)_{5 \alpha}$ for $\alpha = \{ e, \, \mu, \, \tau \}$ (eqn. ~\ref{eq:mass5offdg}).   We are able to find a good solution to atmospheric neutrino oscillations with maximal $\nu_\mu \rightarrow \nu_\tau$ mixing, a solution to solar neutrino oscillations in the SMA MSW region and a fit to LSND.  The fit is presented in tables ~\ref{t:5numass2} and ~\ref{t:5nuangles} and in figures 9(a,b), 10(a,b) and 11.

Note, the parameter $d$ (table ~\ref{t:5numass2} and eqn.~\ref{eq:mass5offdg}) 
$ = \; \frac{\mu^\prime \, V^\prime_{16}}{m_t \, \phi}$.  Thus
$\mu^\prime =   m_t \, d \, \frac{\phi}{V^\prime_{16}} = 0.28 \, \frac{\phi}{V^\prime_{16}} $ GeV.   In addition, we have $ m' =  0.838 $ eV [table ~\ref{t:5numass2}] $\approx {m_t^2 \ \omega \ \phi  \over  V_{16} \ V'_{16}}$.  Thus we find $ V_{16} \sim  {m_t^2 \ \omega \over m'} \frac{\phi}{V^\prime_{16}}  \sim 5.3 \times 10^{13} \, \frac{\phi}{V^\prime_{16}} $ GeV.  In order to obtain this solution without fine tuning we must assume that the ratio  $\frac{\phi}{V^\prime_{16}} \sim 100$.  As in the previous four neutrino case, this may be attributable to ratios of Yukawa couplings.   

\protect
\begin{table}
\caption[3]{
{\bf Fit to atmospheric, solar  and LSND neutrino oscillations 
[5 neutrinos \  SMA MSW + LSND]} \\\
\mbox{Initial parameters: $|\kappa_1| = |\kappa_2|^2, \; |\kappa_2| = 0.032$ }\ \ \ \hspace{1in}

$m' = 0.8380$ eV , \ $b$ = 0.9015, \ $d$ = 0.0016, \ $\Phi_b$ = -3.18rad, \ $\Phi_d$ = -4.83rad 
}
\label{t:5numass2}
$$
\begin{array}{|c|c|}
\hline
{\rm Observable} &{\rm Computed \;\; value} \\
\hline
\delta m^2_{atm}            &  3.7 \cdot 10^{-3} \ \rm eV^2          \\
\sin^2 2\theta_{atm}            &  0.99        \\
 \delta m^2_{sol}   &  5.7 \cdot 10^{-6}  \ \rm eV^2  \\
\sin^2 2\theta_{sol} &  4.0 \cdot 10^{-3}      \\
\delta m^2_{LSND} &  0.36  \\
\sin^2 2\theta_{LSND} &  0.026  \\
\hline
\end{array}$$
\end{table}

\protect
\begin{table}
\caption[3]{
{\bf Neutrino Masses and Mixings [5 neutrinos \  SMA MSW + LSND]} \\

\mbox{Mass eigenvalues [eV]: \ \  $0.88 \cdot 10^{-7}$, \ 0.0007, \ 0.0025, \ 0.6013, \  0.6043 \hspace{1cm}} \\
\mbox{Magnitude of neutrino mixing matrix  U$_{\alpha i}$ \hspace{1.7cm}}\\
\mbox{ $i = 1, \cdots, 5$ -- labels mass eigenstates. \hspace{1.5cm}} \\
\mbox{ $\alpha = \{ e, \ \mu, \ \tau, \ s_1, \ s_2 \}$ labels flavor eigenstates.}
}
\label{t:5nuangles}
$$
\left[ \begin{array}{ccccc}
0.0586       &  0.9940      & 0.0297     & 0.0763 &     0.0430 \\
0.0033      &    0.0802     & 0.0182      & 0.6998 &  0.7096   \\
0.0018   & 0.0356     & 0.0617    & 0.7091 &  0.7015  \\
0.0036  & 0.0291    & 0.9975   &  0.0401  &  0.0507  \\
0.9983  &  0.0585  & 0.0053  & 0.0014 & 0.0015  \\
\end{array} \right]$$
\end{table}

\renewcommand{\thefigure}{9 \alph{figure}}\setcounter{figure}{0}
\begin{figure}
	\centerline{ \psfig{file=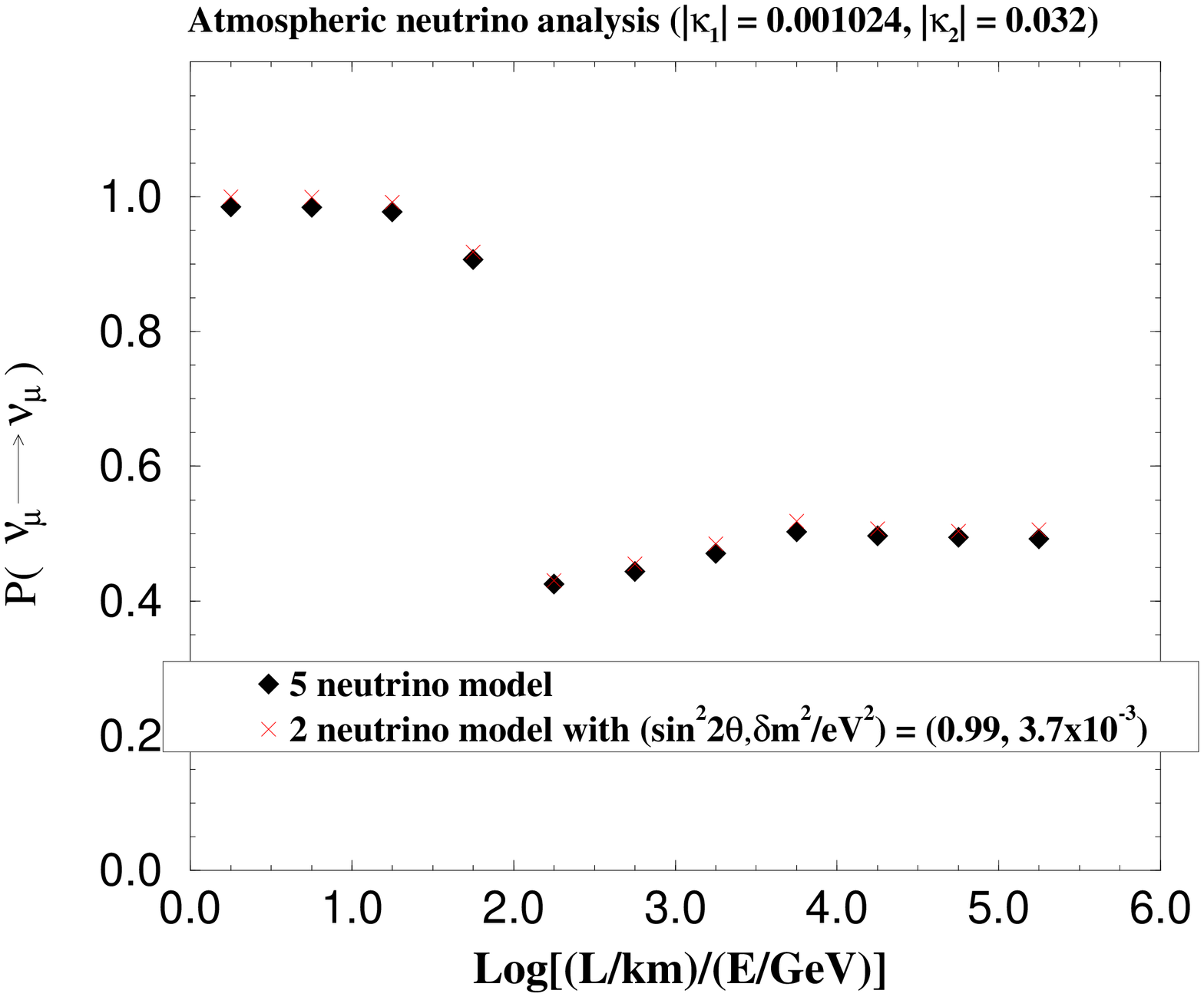,width=9cm,angle=-90}}
\caption{Probability $P(\nu_\mu \longrightarrow \nu_\mu)$ for atmospheric neutrinos.  For this analysis, we neglect matter effects.}
\end{figure}
\begin{figure}
	\centerline{ \psfig{file=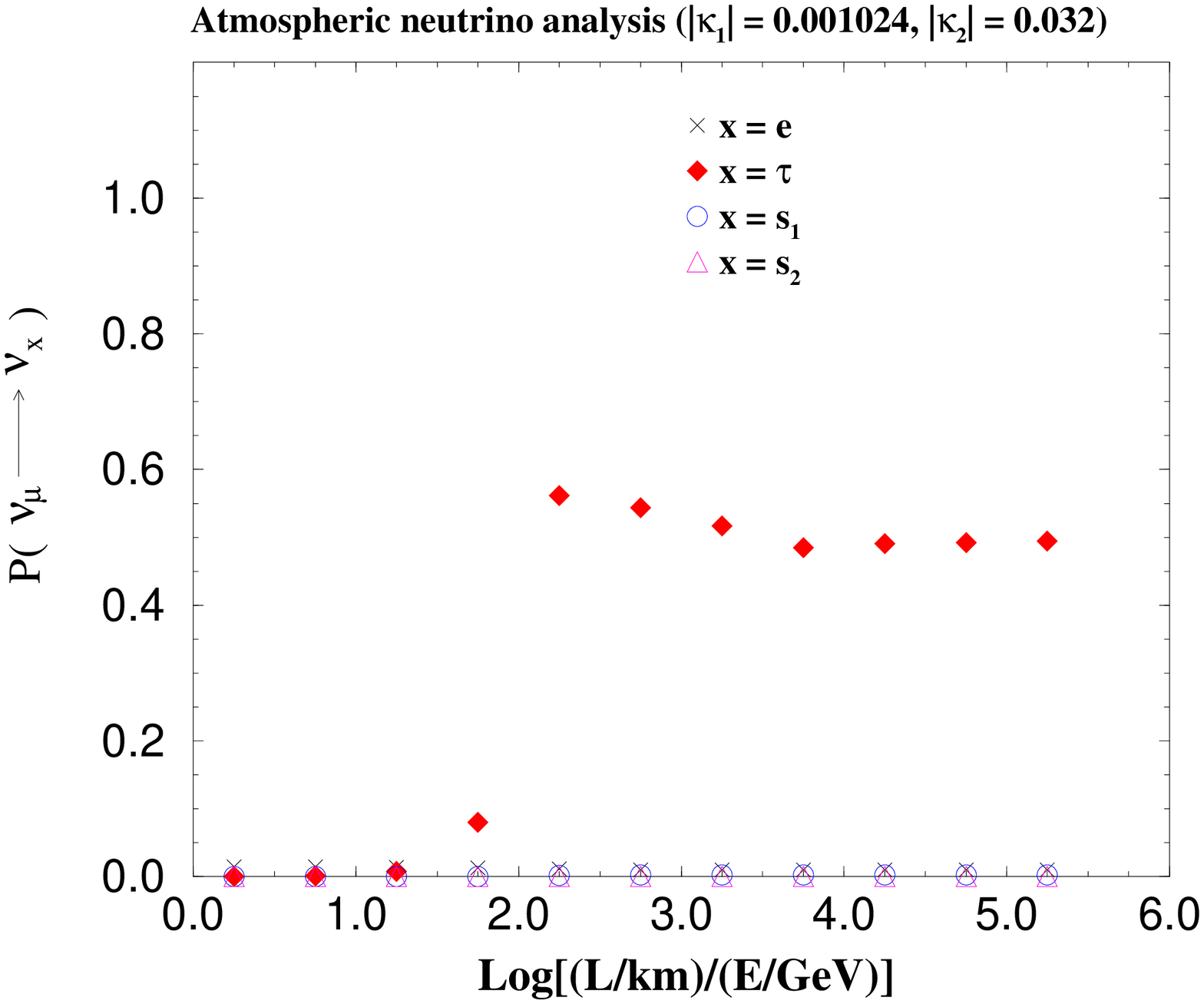,width=9cm,angle=-90}}
\caption{Probabilities $P(\nu_\mu \longrightarrow \nu_x)$ ($x=e$, $\tau$, $s_1$ and $s_2$) for atmospheric neutrinos.}
\end{figure}

\renewcommand{\thefigure}{10 \alph{figure}}\setcounter{figure}{0}
\begin{figure}
	\centerline{ \psfig{file=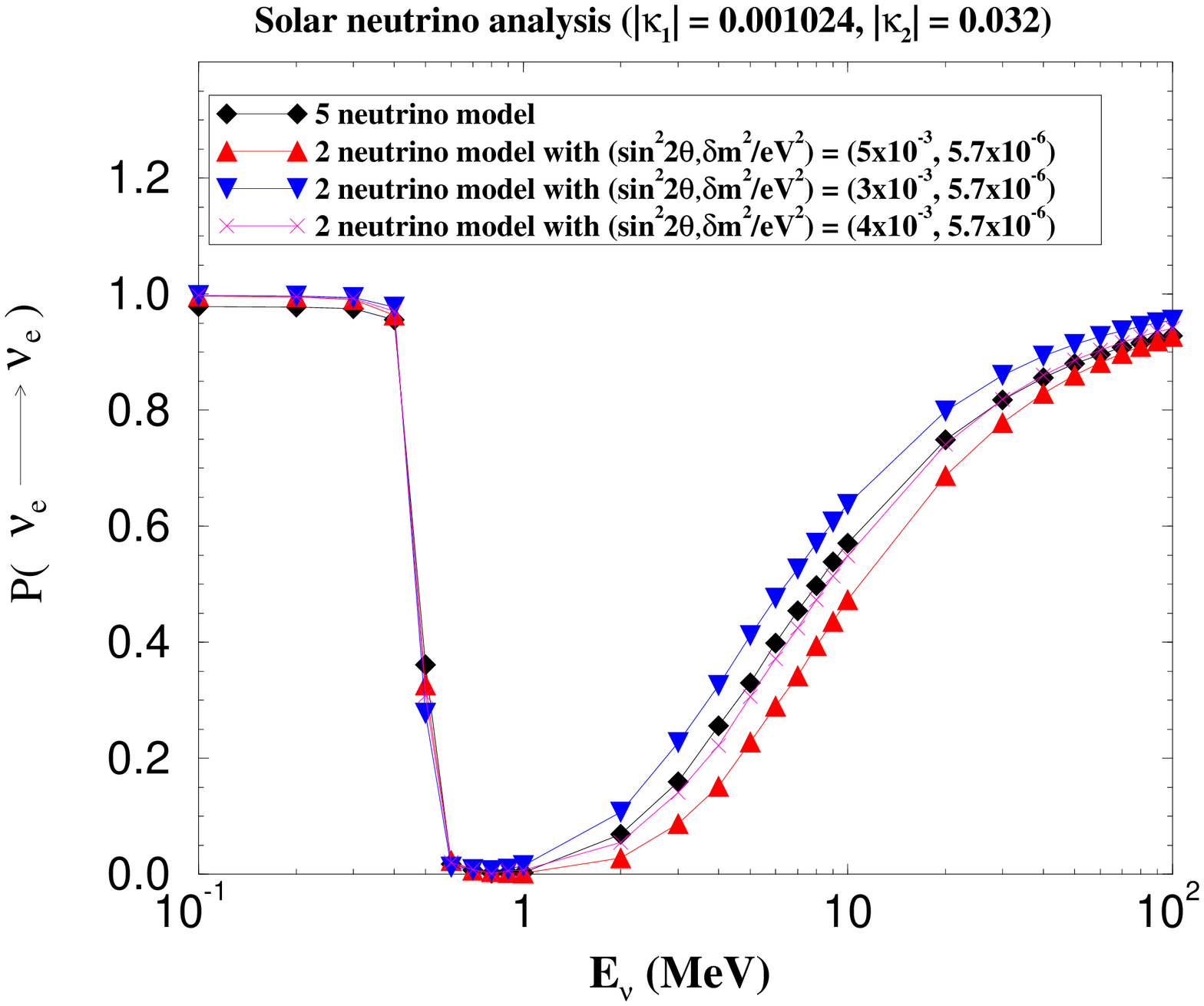,width=9cm,angle=-90}}
\caption{Probability $P(\nu_e \longrightarrow \nu_e)$ for solar neutrinos.}
\end{figure}
\begin{figure}
	\centerline{ \psfig{file=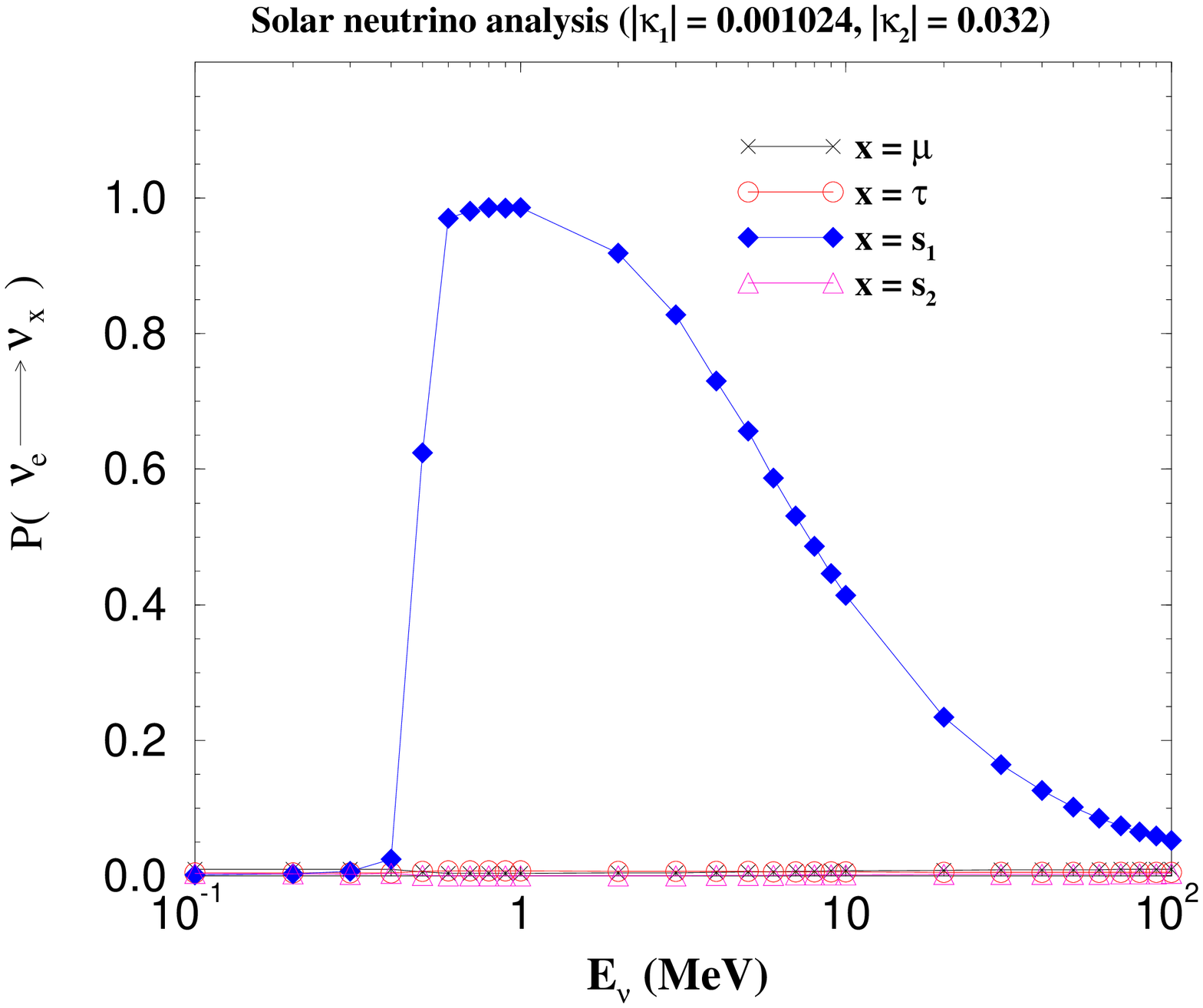,width=9cm,angle=-90}}
\caption{Probabilities $P(\nu_e \longrightarrow \nu_x)$ ($x=\mu$, $\tau$, $s_1$ and $s_2$) for solar neutrinos.}
\end{figure}
\renewcommand{\thefigure}{11}\setcounter{figure}{0}
\begin{figure}
	\centerline{ \psfig{file=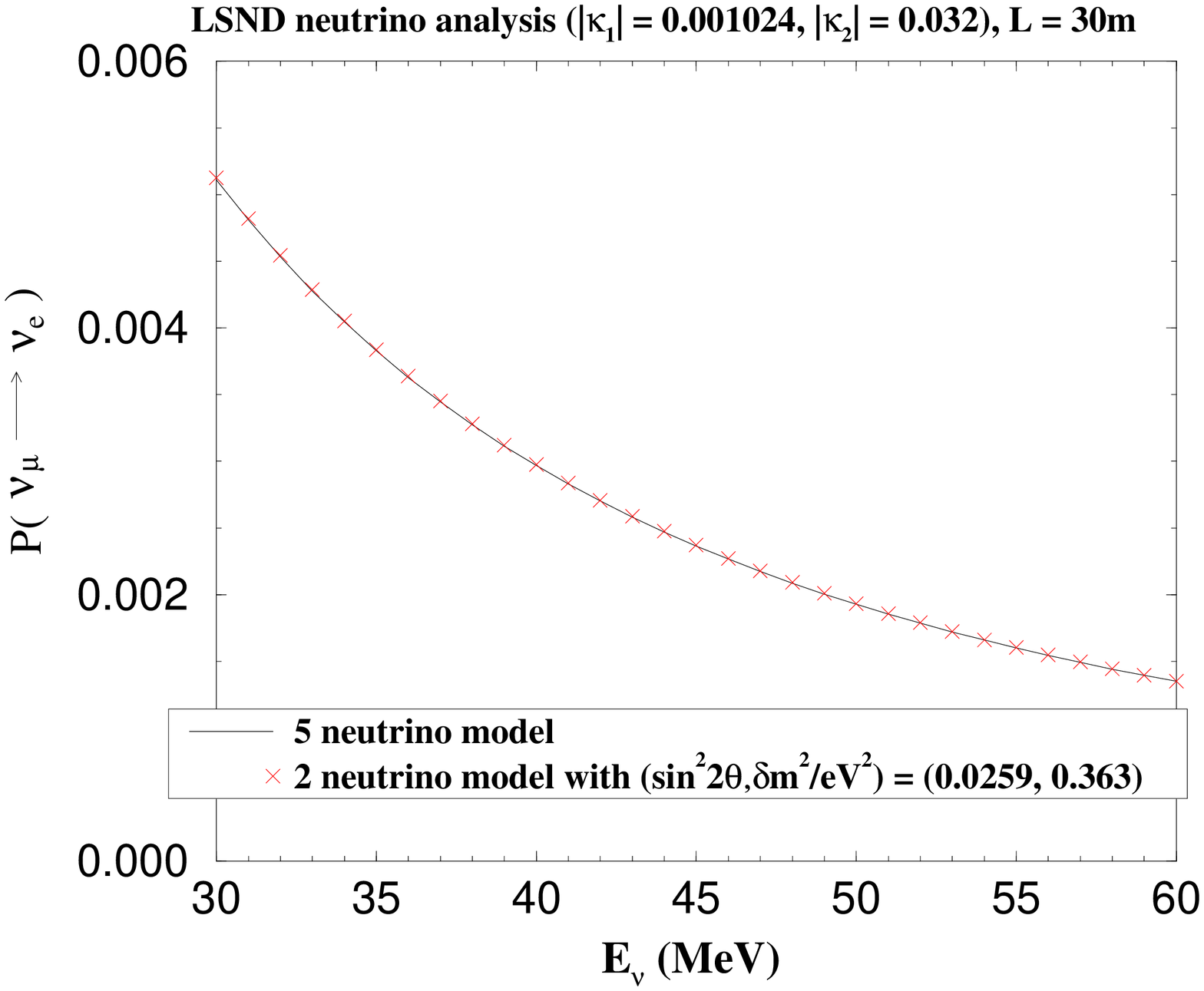,width=9cm,angle=-90}}
\caption{Probability $P(\nu_\mu \longrightarrow \nu_e)$ for LSND energies.}
\end{figure}

\section{ Discussion }

In this paper we analyze the predictions for both charged fermion {\em and} neutrino masses and mixing angles in an SO(10) SUSY GUT with U(2)$ \times$U(1)$^n$ family symmetry.   We find that, if we allow for the most general family symmetry breaking vevs, the model can accommodate three different three-neutrino oscillation solutions to atmospheric and solar neutrino data, one four and one five neutrino solution to atmospheric, solar and LSND data.  We also find a three neutrino solution to atmospheric and LSND data alone.  In spite of all this freedom in the neutrino sector, the fits for charged fermion masses and mixing angles are relatively unaffected.  

In all cases we find atmospheric neutrino data described by maximal $\nu_\mu \rightarrow \nu_\tau$ mixing.\footnote{We have not searched for solutions with   
maximal $\nu_\mu \rightarrow \nu_s$ mixing, since this is not favored by the latest Super-Kamiokande data~\cite{atmos}.} Super-Kamiokande is able to distinguish $\nu_\mu \rightarrow \nu_x$ for $\nu_x = \nu_\tau$ or $\nu_s$ (see talks by K. Scholberg and W.A. Mann~\cite{atmos}).   There are two proposed methods.  The first uses the measured zenith angle dependence, since there is an MSW effect in the earth for  $\nu_s$ but not for $\nu_\tau$.  This effect suppresses $\nu_\mu \rightarrow \nu_s$ oscillations for high energy neutrinos coming from below.  Recent data does not show such an effect; thus favoring $\nu_x = \nu_\tau$.  The second method uses the ratio of neutral current [NC] to charged current [CC] processes which can distinguish between the two.   Here there is preliminary data favoring $\nu_x = \nu_\tau$.   This ratio satisfies
 \begin{eqnarray}
R_{(NC/CC)}  & < 1 & \;\; \rm for \; \nu_x = \nu_s  \label{eq:R1} \\ 
& = 1  & \;\; \rm for \; \nu_x = \nu_\tau  . \nonumber 
\end{eqnarray} 
Using SuperK data for $\pi^0$ events produced by neutral current neutrino scattering in the detector one measures

\begin{eqnarray}
R_{(NC/CC)}  & \equiv  \frac{(\pi^0/e)_{Data}}{(\pi^0/e)_{MonteCarlo}} & \label{eq:R2} \\
&  =  1.11  \pm  0.06 ({\rm data \; stat.}) \pm 0.02 ({\rm MC \; stat.}) \pm 0.26 ({\rm sys.}) & \nonumber 
\end{eqnarray} 

The oscillations $\nu_\mu \rightarrow \nu_\tau$  may also be visible at long baseline neutrino experiments.  Both K2K ~\cite{k2k} and MINOS ~\cite{minos} are designed to test for $\nu_\mu$ disappearance.  For example at K2K~\cite{k2k},  the mean neutrino energy $E = 1.4 $GeV and distance $L = 250$ km corresponds to a value of x = 2.3 (see figures 2a, 4a, 6a and 9a) and hence  $P(\nu_\mu \rightarrow \nu_\mu) \sim .45$.  

Results on solar neutrino oscillations or LSND will, on the other hand, be able to narrow down the acceptable regions of parameter space, but cannot test this class of models.

Finally, the plethora of solutions presented in this paper is in stark contrast to the unique solution obtained assuming the minimal family symmetry breaking vevs studied previously in paper I ~\cite{brt}. In the latter case we cannot find any three family solutions to both atmospheric and solar data and we find a unique four neutrino solution to atmospheric and solar data but NOT LSND.  Thus it is clear that the neutrino sector is in general much less constrained than charged fermions.  Nevertheless, it is pleasing to find a simple SUSY GUT which can accommodate all of this low energy data.   

\noindent
{\bf Acknowledgements}
This work is partially supported by DOE grant DOE/ER/01545-767.

\vfill

\section{Appendix}

\noindent
{\bf Solar neutrino analysis}

In this appendix we describe in detail the approximation which we used in the numerical analysis of solar neutrino oscillations.
The Schr\"{o}dinger equation for solar neutrinos is given by
\begin{eqnarray}
i\frac{d}{dt} \Psi_{\nu}^\alpha(t) &=& H_{\alpha \beta} \Psi_\nu^\beta(t), \\
H_{\alpha \beta}&=& \frac{(m_\nu^\dagger m_\nu)_{\alpha \beta}}{2E} 
+ V_{\alpha}(t) \delta_{\alpha \beta}.
\end{eqnarray}
Here $\Psi_\nu^\alpha$ is a state vector for neutrinos with flavor $\alpha$
($\alpha=e$, $\mu$, $\tau$, and $s$ for four neutrino model
\footnote{Here we present our method of solar neutrino analysis
in a four neutrino model. The method can be easily extended to a three 
neutrino model or a model with more neutrinos.}),  $H$ is the Hamiltonian for solar neutrinos, and $E$ is the neutrino energy. The mass matrix $ m_\nu $ in the flavor basis is given by (see equation ~\ref{eq:mnu})
\begin{eqnarray}
m_\nu^{diag} = & U^T m_\nu U &,
\end{eqnarray}
where $U$ is the mixing matrix for neutrinos 
($\nu^{\rm flavor}_\alpha=U_{\alpha i} \; \nu^{\rm mass}_i$ where
$i=1-4$  for four neutrino model)
and $ m_\nu^{diag}$ is the diagonal mass matrix in the mass eigenstate basis.
$V_\alpha$(t) is a time-dependent potential for neutrinos with 
flavor $\alpha$ as follows:
\begin{eqnarray}
V_{e}(t) &=& \sqrt{2} G_F \{n_e(t) - \frac{1}{2} n_n(t)\},
\nonumber \\
V_{\mu}(t) &=& V_{\tau}(t) = - \sqrt{2} G_F \frac{1}{2} n_n(t),
\nonumber \\
V_{s}(t) &=& 0,
\end{eqnarray}
where $G_F$ is the Fermi coupling constant. Here we assume that
electron ($n_e$) and neutron ($n_n$) number densities at a distance 
$r=ct$ from the center of the sun are given by
\begin{eqnarray}
n_e &=& 4.6 \times 10^{11} \exp\left(-10.5 \frac{r}{R}
\right)~{\rm eV}^3,
\\
n_n &=& 2.2 \times 10^{11} \exp\left(-10.5 \frac{r}{R}
\right)~{\rm eV}^3,
\end{eqnarray}
where $R$ is a solar radius.

Mass scales for the atmospheric and LSND neutrino problems ($\delta m^2_{\rm atm.} \simeq 10^{-3}$ eV$^2$, $\delta m^2_{\rm LSND} \simeq 1$ eV$^2$) are much larger that for the solar neutrino problem ($\delta m^2_{\rm solar} \leq 10^{-5}$ eV$^2$).  When we include the mass scales for atmospheric and/or LSND neutrinos and solve the Schr\"odinger equation for the solar neutrino problem, it is almost impossible to solve it numerically because of these larger mass scales and the rapid fluctuations they produce.  Thus, in order to solve the Schr\"odinger equation numerically, we use the following approximation.

We divide the mass term $ m_\nu^\dagger m_\nu $ into two parts:
\begin{eqnarray}
m_\nu^\dagger m_\nu &=& U \; (m_\nu^{diag})^\dagger m_\nu^{diag} \; U^\dagger,
\nonumber \\
        &=& U \; m_L^2 \; U^\dagger + U \; m_H^2 \; U^\dagger,
\end{eqnarray}
where $m_L^2$ ($m_H^2$) is a ``Light'' (``Heavy'') part,
\begin{eqnarray}
m^2_L &=& \left(
\begin{array}{cccc}
m_1^2 & & & \\
 & m_2^2 & & \\
 & & 0 & \\
 & & & 0 \\
\end{array}
\right),
\\
m_H^2 &=& \left(
\begin{array}{cccc}
0 & & & \\
 & 0 & & \\
 & & m_3^2 & \\
 & & & m_4^2 \\
\end{array}
\right),
\end{eqnarray}
and we assume that $\delta m^2_{21}=m^2_2-m^2_1$ is the scale for 
solar neutrino problem and $ m_1^2 < m_2^2  \ll m_3^2 < m_4^2$.
Then the Hamiltonian $H$ is given as
\begin{eqnarray}
H &=& H_H + H_L, \nonumber \\
H_H &=& U \frac{m_H^2}{2E} U^\dagger, \\
H_L &=& U \frac{m_L^2}{2E} U^\dagger + V_\alpha(t) \delta_{\alpha \beta}.
\end{eqnarray}

The state vector is also divided into two parts as follows:
\begin{eqnarray}
\Psi_\nu^\alpha(t) = A_{\alpha \beta}(t) \; \Phi_\nu^\beta(t),
\label{state_nu}
\end{eqnarray}
where we define $A$ to satisfy the following equation:
\begin{eqnarray}
i \frac{d}{dt} A(t) &=& H_H A(t), \nonumber \\
 A(t=0) &=& I
\label{eq_for_A}
\end{eqnarray}
where $I$ is a unit matrix. We can easily solve the equation (~\ref{eq_for_A})
and the solution is given by
\begin{eqnarray}
A(t) &=& \exp (-i H_H t), \nonumber \\
  &=& U \exp \left(-i \frac{m_H^2}{2E}t \right) U^\dagger.
\end{eqnarray}
Then $\Phi_\nu$ satisfies
\begin{eqnarray}
i \frac{d}{dt} \Phi_\nu(t) = U 
\left[ 
\frac{m_L^2}{2E} + \exp \left( i \frac{m_H^2}{2E} t \right) 
U^\dagger V(t) U 
\exp \left( -i \frac{m_H^2}{2E} t \right) \right] U^\dagger \Phi_\nu(t).
\label{eq_for_phi}
\end{eqnarray}

Since the mass scales $m_{3,4}$ included in the matrix $m_H^2$ 
are too large for MSW effects, the exponential terms 
$\exp \left( \pm i \frac{m_{3,4}^2}{2E} t \right)$ oscillate rapidly.
Therefore we replace them by their time-averaged values:
\begin{eqnarray}
\exp \left( \pm i \frac{m_{3,4}^2}{2E} t \right)
&\rightarrow& 0
\end{eqnarray}
Then equation \ref{eq_for_phi} has the following approximate form
\begin{eqnarray}
i \frac{d}{dt} \Phi_\nu^\alpha(t) &\simeq& \left[
U_{\alpha i} \left(
\frac{m^2_{L i}}{2E} \delta_{ij} + U^\dagger_{i \gamma} 
V_\gamma(t) U_{\gamma j}
\right)
U^\dagger_{j \beta} \right. \nonumber \\
&& \left.+U_{\alpha~ i+2} U^\dagger_{i+2~ \gamma} V_\gamma(t) U_{\gamma~ i+2} 
U^\dagger_{i+2~ \beta}
\right]
\Phi_\nu^\beta(t),
\label{schrodinger_eq}
\end{eqnarray}
where the indices $\alpha,~\beta,~\gamma$ run from 1 to 4, on the other hand,
the indices $i,~j$ from 1 to 2.
We then solve equation ~\ref{schrodinger_eq} with the initial condition
\begin{eqnarray}
\Phi_\nu(t=0)=(1,0,0,0)~~~{\rm or}~~~ \Psi_\nu(t=0)=(1,0,0,0) .
\end{eqnarray}

Finally, the oscillation  probability $P(\nu_e \rightarrow \nu_\alpha)$ 
($\alpha=e,\mu,\tau$ or $s$) at time $t$ is given by
\begin{eqnarray}
P(\nu_e \rightarrow \nu_\alpha) &=& \left| \Psi_\nu^\alpha(t) \right|^2
\nonumber \\
&=& \left| A_{\alpha \beta} (t) \; \Phi_\nu^\beta(t) \right|^2
\nonumber \\
&\simeq& \left| \sum_{i=1,2,\beta=1-4} 
U_{\alpha i} \ U^\dagger_{i \beta} \ \Phi_\nu^\beta \right|^2
+ \sum_{i=3,4} \left| \sum_{\beta=1-4} 
U_{\alpha i} \ U^\dagger_{i \beta} \ \Phi_\nu^\beta \right|^2.
\label{prob}
\end{eqnarray}
where the $\simeq$ in the last line (equation \ref{prob}) refers to the fact that the time average of $\exp(\pm i \frac{m_{3,4}^2}{2E}t)$ was used.

%

\end{document}